\documentclass[]{aa}

\newcommand{\nc}{\newcommand}

\newcommand{\HII}{H {\sc ii}}
\nc{\msun}{\ensuremath{\mathrm{M}_\odot}}
\nc{\lsun}{\ensuremath{\mathrm{L}_\odot}}
\nc{\thCO}{$^{13}$CO}
\nc{\CeiO}{C$^{18}$O}
\nc{\ntwhp}{N$_2$H$^+$}
\nc{\kms}{\mbox{km~s$^{-1}$}}
\nc{\Kkms}{\mbox{K\,km~s$^{-1}$}}
\nc{\twCO}{$^{12}$CO}
\nc{\hcop}{HCO$^+$}
\nc{\htwoco}{H$_2$CO}
\nc{\chthrcn}{CH$_3$CN}
\nc{\chthroh}{CH$_3$OH}
\nc{\hcthn}{HC$_3$N}
\RequirePackage{etex}
\renewcommand\arcmin{\mbox{$^\prime$}}%
\renewcommand\arcsec{\mbox{$^{\prime\prime}$}}%
\nc{\cmsq}{\mbox{cm$^{-2}$}}
\nc{\cmcub}{\mbox{cm$^{-3}$}}
\nc{\permillionyear}{Myr$^{-1}$}

\nc{\vlsr}{v$_{\rm lsr}$}
\nc{\coldens}{N$_{col}$}
\nc{\chisq}{$\chi^2$}
\nc\micron{\mbox{$\mu$m}}
\newcommand\arcdeg{\mbox{$^\circ$}}%

\newcommand\phn{\phantom{0}}%

\usepackage{graphicx}	
\usepackage{amsmath}    
\usepackage{subcaption}
\usepackage{enumerate}
\usepackage{graphicx}
\usepackage{txfonts}
\usepackage{lipsum}

\usepackage{lscape}             
\usepackage{placeins} 

\usepackage{url}
\usepackage{hyperref}
\hypersetup{
    colorlinks=true,
    citecolor=blue,
    linkcolor=blue,
    filecolor=blue,      
    urlcolor=blue,
    pdftitle={G183}
    }

\begin{document}



\title{G183: An outer galaxy filament feeding a massive protostar}
\titlerunning{High-mass star formation in G183}


\author{Bhaswati Mookerjea\inst{1}
\and
Saurav Sen\inst{1}
\and
V. S. Veena\inst{2}
\and
Carsten Kramer\inst{3}}
\institute{Department of Astronomy \& Astrophysics, Tata Institute of
Fundamental Research,\\ Homi Bhabha Road, Mumbai 400005, India
\email{bhaswati@tifr.res.in}
\and
Max Planck Institut f\"ur Radioastronomie, Auf dem H\"ugel 69, D-53121 Bonn, Germany
\and
Institut de Radioastronomie Millim\'etrique (IRAM), 300 rue de la Piscine, 38400 Saint-Martin-d’H\'eres, France}


\abstract
 {We present the first detailed multi-tracer observation of a 5-pc long
 outer Galaxy filament, G183, and the massive young stellar object (YSO)
 IRAS\,5480+2545 associated with it. Using the IRAM 30-m telescope at
 $\lambda = $1.4 and 3\,mm, we probed the molecular gas distribution at
 angular resolutions  of $\sim 12$\arcsec $-28$\arcsec (0.1--0.3\,pc at
 $d = 2.1$\,kpc). The velocity-resolved \CeiO(1--0) observations
 conclusively show a main filament with a skeleton of ridges.  The main
 filament is a 5\,pc long velocity-coherent structure with a continuous
 and quiescent velocity field along its length up to the star-forming
 hub that accretes mass from the filament. The internal gas kinematics
 of most of the G183 filament is dominated by thermal motions
 ($\sigma_{\rm NT}$/c$_s\sim 1$) and large-scale velocity gradients
 arising due to outflows and accretion of matter in the massive YSO. The
 dispersion-size relation almost up to 1\,pc is consistent with Larson's
 law, suggesting that the origin of the filament is a turbulence
 cascade. The massive YSO, S1, with no corresponding radio continuum
 detection is characterized as a high-mass protostellar object with a
 mass of 156\,\msun\ and an $M/L$ ratio of 0.04. We identify a kinematic
 signature of the accretion of material from the filament onto the YSO,
 S1. The rates of molecular gas accretion and entrainment in S1 are
 estimated to be 8.6 and 2.6 (in units of 10$^{-4}$\,\msun\,yr$^{-1}$),
 respectively. In comparison to the inner Galaxy high-mass star-forming
 filaments forming massive stars, G183 has a lower column density;
 however, the accretion and outflow rates in S1 are similar. The
 detection of hydrocarbons such as \chthrcn\ and \hcthn\  indicates the presence of hot-core chemistry in S1. These results highlight the universality of physical processes involved in massive star formation across a range of Galactic environments.}



\maketitle


\section{Introduction}

Improved observing facilities at long wavelengths (infrared and beyond) such as Spitzer and Herschel have led to the understanding that the molecular interstellar medium is preferentially organized in filaments \citep[e.g.,][]{Schneider1979, Myers2009,Andre2010, Molinari2010, Peretto2014}) and that dense cores in them are the sites of star formation.  Filaments are
ubiquitous in the Galaxy and are observed over a large range of scales. The physics of the filaments being apparently tightly connected to star formation has motivated in-depth studies to explain the origin and evolution of these structures.  Filamentary structures pervading clouds are unstable against both radial collapse and fragmentation \citep[e.g.,][]{Larson1985,
Inutsuka1997}, and although their origin or formation process is still unclear, turbulence and gravity \citep[e.g.,][]{Klessen2000, Andre2010} can produce, together with the presence of magnetic fields \citep[e.g.,][]{Molina2012, Kirk2015}, the observed structures. It is thought that star formation occurs preferentially along the filaments, with high-mass stars forming in the highest-density regions, called ridges or hubs \citep[$N_{\rm H}\sim 10^{23}$\,\cmsq\ and $n_{\rm H_2}\sim 10^6$\,\cmcub, e.g.,][]{Schneider2010, Peretto2013, Peretto2014}, where several filaments converge.  This scenario also constitutes the backbone of one of the competing theories of high-mass star formation, global hierarchical collapse (GHC), in which all size scales contract gravitationally, and accrete from the next larger scale \citep{Vazquez2019}. 
In the recent years, an increasing number of works have focused on the study of the dynamics and fragmentation of filamentary structures from both the observational and theoretical points of view \citep[see e.g.,][]{Andre2010, Arzoumanian2019, Williams2018, Clarke2019}. 

Observational studies of formation of massive stars is hindered by the facts that high-mass stars are fewer in number, and hence typically at larger distances and the protostars being deeply embedded in dusty clouds require high-angular-resolution long-wavelength ($>100$\,\micron) observations. While the body of work on star formation in filaments and hub-filament systems (HFSs) is growing, most observations are directed towards the inner Galaxy, which is both structurally and kinematically complex due to the presence of multiple spiral arms as well as regions with multiple outflows from embedded protostars along the line of sight. The galactocentric distribution reveals that compared to the inner Galaxy, surface density of the high-mass star-forming regions  and the proportion of clumps associated with methanol-masers (an indicator of high-mass star formation) is significantly lower in the outer Galaxy  \citep{urquhart_2024}.  The outer Galaxy with a lower metallicity \citep{Rudolph1997}, lower average temperature in molecular clouds \citep{Mead1988}, higher gas-to-dust ratio \citep{Gianetti2017b}, and lower gamma-ray flux \citep{Bloemen1984} is expected to have a rather different star-forming environment. Since the first series of pointed observations of IRAS sources \citep{Brand1986, Wouterloot1988}, there have only been two major molecular line surveys in the outer Galaxy. Of these recent surveys, the Forgotten Quadrant Survey (FQS) observed CO and \thCO(1--0) with a beam size of 55\arcsec\ \citep{Benedittini2020} and the Outer Galaxy High Resolution Survey (OGHRES) observed  CO(2--1) at a resolution of 25\arcsec \citep{Colombo2021,urquhart_2024,Urquhart2025}. While high-mass star formation in the outer Galaxy may be rare, using ALMA observations the first hot molecular core  was recently detected in the extreme outer Galaxy at a galactocentric distance of 19\,kpc \citep{Shimonishi2021}. More recently, the first detection of spatially resolved protostellar outflows and jets in outer Galaxy was reported in intermediate- to high-mass star-forming cores \citep{Ikeda2025}.

We present here the mapping observations of a long ($\sim 5$\,pc) filament G183 located in the outer Galaxy, which has the appearance of a HFS.  It was identified as a high-mass star-forming region (IRAS 05480+2545) based on its IRAS colours and  association with 6.7\,GHz Class II methanol masers \citep{henning_1992, kawamura_1998, klein_2005, wu_2010, wu_2011}. Using multiple molecular line emission maps at 1.4 and 3\,mm observed with the IRAM 30\,m telescope in combination with existing archival data of far-infrared and sub-millimetre dust continuum observations, we investigated the origin of the filament and constrained the physical properties and evolutionary stage  of the massive young stellar object (YSO) associated with G183.

\section{Observation and dataset}
\subsection{New molecular line observations with IRAM 30m}

A 10\arcmin$\times$10\arcmin\  region around G183 centred at R.A. = 05:51:10.90 Dec = +25:46:14.7 (2000) was observed with the IRAM 30 m telescope in September and October 2022 with the EMIR receiver and the Fourier Transform Spectrometer (FTS) backends at 3 mm as part of the project 055-22. The receiver was tuned to a central frequency of 100.9 GHz with dual-polarization to cover the transitions of large-scale molecular gas tracers \thCO(1--0), \CeiO(1--0), the dense gas tracers \hcop(1--0), HCN(1–0), HNC(1–0), and the cold and dense gas tracer \ntwhp(1--0). This frequency setting also covered the \hcthn(10--9), \hcthn(12--11) and \chthrcn\ (5--4) and (6--5) transitions. The FTS backends were used to cover the 16 GHz bandwidth of the receiver with a uniform spectral resolution of 200 kHz, which results in a velocity resolution of $0.65$\,\kms\ at 3\,mm. The observations were carried out in the on-the-fly (OTF) mode employing position switching to an OFF position sufficiently far away. The map was observed as a mosaic of boxes each 2\farcm5$\times$2\farcm5. The beam sizes for the 3\,mm setting range between 22\farcs3--27\farcs8. 

We also observed 4\farcm2$\times$4\farcm2 maps of G183 at frequencies between 215--230\,GHz (1.4\,mm) in January-February 2025 as part of the project 147-24. The EMIR receiver was tuned to 234.45\,GHz in the outer USB and FTS was used as the backend with a velocity resolution of 0.25\,\kms. This frequency setting covered the large-scale molecular gas tracers CO, \thCO\ , and \CeiO\ (2--1) and the outflow tracers SO($5_6$--$4_5$), SO($5_5$--$4_4$), and \htwoco(3--2). The beam sizes for the spectral lines observed with this setting range between 10\farcs7--11\arcsec. 

The spectra were calibrated with CLASS, which is part of the GILDAS\footnote{https://www.iram.fr/IRAMFR/GILDAS} software package. The data were converted from the antenna temperature ($T_{\rm A}^\ast$) to the main beam brightness temperature ($T_{\rm mb}$) using a forward efficiency ($F_{\rm eff}$ = 95\% at 3 mm and 92\% at 1.4\,mm), a main beam efficiency (0.79--0.81 at 3\,mm and 0.58-0.61 at 1.4\,mm) within {\sc CLASS} estimated  by using Ruze's formula and the equation $T_{\rm mb}$ = $T_{\rm A}^\ast\times F_{\rm eff}/B_{\rm eff}$.  A linear function was fitted to the line free channels of the spectra to subtract the baseline. Finally 3 mm data were all smoothed to a common spectral resolution of 0.7\,\kms\ , resulting in a median root-mean-square (rms) of 8\,mK. The 1.4\,mm data were all smoothed to a common velocity resolution of 0.3\,\kms\ leading to a median rms of 0.14\,K.

\subsection{Archival data: CO(3--2) and infrared continuum}

A 7\arcmin$\times$7\arcmin\ CO(3--2) map of the G183 region centred around the massive star-forming clump was retrieved from the archives of the James Clerk Maxwell Telescope (JCMT) for comparison with the low-frequency data from IRAM 30m. The beam size of the CO(3--2) map is 15\arcsec. We also used dust continuum images of the region observed with the PACS and SPIRE instruments on board Herschel as well as a 350\,\micron\ image at 8\farcs5 resolution obtained as a part of the Bolocam Galactic Plane Survey (BGPS). All the far-infrared and sub-millimetre continuum images were downloaded from the NASA/IPAC Infrared Science Archive (IRSA).

\section{Overview of the source
        \label{section_3}}
\begin{figure*}
    \centering
   \includegraphics[width=0.95\linewidth]{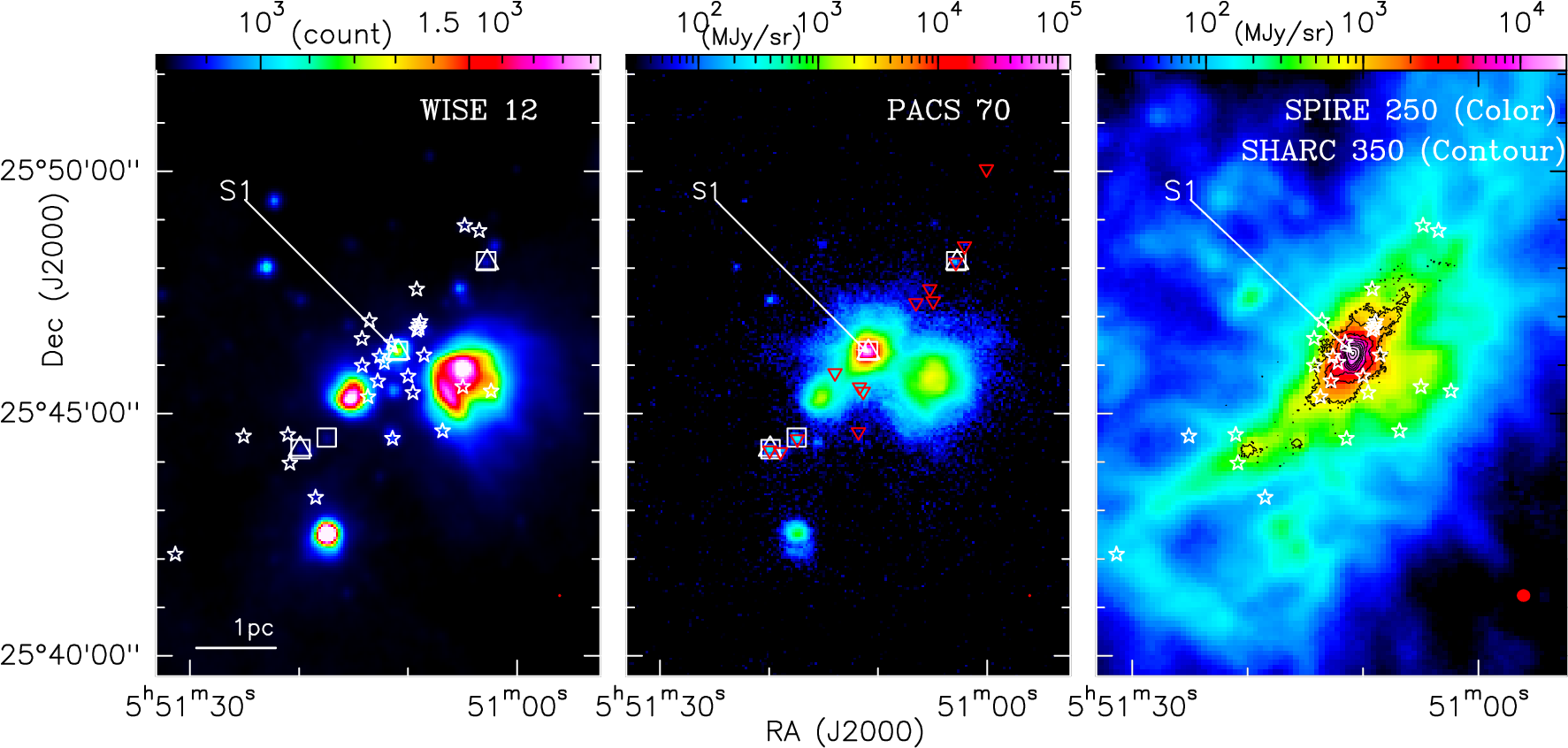}
    \caption{Continuum images of the G183 region. {\em Left} WISE 12\,\micron\ . {\em Middle} Herschel/PACS at 70\,\micron\ . ({\em Right:}) 250\,\micron\ SPIRE image (colour) with 350\,\micron\ SHARC image with a beam size of 8\farcs5 as contours. Contour levels (Jy/beam) are 0.5, 2, 4, 6, 8, 10, 12 to 30 in steps of 8. Continuum sources at 70 and 160\,\micron\ shown as white open squares and triangles, respectively. Class 1 and Class 2 sources from GLIMPSE360 are shown as inverted red triangles and white asterisks. The location of the massive YSO S1 is also marked for reference.         \label{fig_g183cont}}
\end{figure*}

The source G183 ($l$=183.348620\arcdeg\ $b$=-0.575879\arcdeg) was identified as a HFS based on Herschel/SPIRE 250\,\micron\ maps of continuum emission \citep{kumar_2022}. The infrared source in the hub of the HFS was detected in the MSX survey and was identified as a massive YSO. Using kinematic information \citet{Mottram2011} estimated a distance of 2.1\,kpc and a bolometric luminosity  ($L_{\rm bol}$) of  4170\,\lsun\ for the source based on the Herschel data. At 12 and 70\,\micron, the emission from the region  primarily reveal the distribution of hot dust, which is distinctly different from the high-column-density cold dust detected at wavelengths of 250\,\micron\ and longer (Fig.\,\ref{fig_g183cont}).  The cold dust emission is extended over 750\arcsec\ (7.6\,pc at 2.1\,kpc) diagonally across the map, with multiple fainter filament-like features joining along its length. The 70 and 160\,\micron\ emission have a common peak (marked as S1 in Fig.\,\ref{fig_g183cont}) that coincides with the source IRAS 05480+2545 and the associated methanol and OH masers \citep{slysh_1997,slysh_1999}. The detection of Class II methanol and the OH masers  towards this IRAS source is indicative of the presence of a massive YSO associated with IRAS 05480+2545. In addition to the elongated feature, the 250\,\micron\ emission also shows diffuse emission from the region. A second peak lying nearly to the west of the peak corresponding to S1 is significantly brighter at 12 and 70\,\micron\ and is not detected at all in the 350\,\micron\ CSO/SHARC II image. The cold dust traced by the emission at 250 and 350\,\micron\ is distributed in the shape of a long filament, with one major condensation associated with the IRAS source 05480+2545 and a few smaller peaks towards the southern part. We performed pixel-by-pixel grey-body fitting of 160, 250, 350, and 500\,\micron\  dust continuum emission using {\em hires}, an improved algorithm for the derivation of high-resolution (18\arcsec) surface densities from multi-wavelength far-infrared Herschel images \citep{Menshchikov2021}. The resulting dust temperature map  shows temperatures peaking at $\sim 20$\,K in the immediate vicinity of the massive YSO S1 but rapidly dropping to temperatures of 11\,K to 15\,K in the filament (Fig.\,\ref{fig_dcolden}). The column density map resembles the 250\,\micron\ cold dust emission with values of $N$(H$_2$) between 0.2--1.7$\times 10^{23}$\,\cmsq\ (Fig.\,\ref{fig_dcolden}). The class I and II sources \citep{winston_2020} identified in the GLIMPSE360 images appear to be distributed along the elongated filament and do not overlap much with the hot dust emission, but rather lie closer to the hub. For the rest of the paper, we thus associate only the cold dust component with the source of our interest and do not consider the western source detected at $\lambda <100$\,\micron\ to be a part of the G183 region.  At a resolution of 8\farcs5, the SHARC II 350\,\micron\ identified a compact clump of 21\arcsec $\times$ 11\arcsec\ size. Based on NANTEN \thCO(1--0) map at a resolution of 2\farcm7 resolution, a molecular cloud at \vlsr = -9.1\,\kms\ with a mass of $\sim$ 300\,\msun\ was identified \citep{kawamura_1998}.  In a recent survey of radio jets from massive protostars, two point-like C-band sources were identified in this region, of these the first one (A) is at the location of the massive YSO S1, a possible radio jet and the other one (B) lies to north-east of (A) offset by more than 40\arcsec\ from S1 \citep{Purser2021}. Since G183 is located at a distance of 2.1\,kpc from us, the metallicity of the source is estimated to be only 20\% lower than the solar luminosity \citep{LuckLambert2011}.

\section{Results}

Here we present results of our mapping observations of molecular line emission at 1.4 and 3\,mm at resolutions of 0.1--0.3\,pc. The lines of CO (and its isotopologues) are useful to trace the distribution of the bulk of the molecular gas. Due to its lower abundance, \CeiO\ is optically thin, and hence serve as an effective tracer of the high column density filamentary structures. Among the other molecules, \hcop, HCN, and HNC are useful tracers of dense gas, although it is possible that at intensity levels of $>1$\,K\,\kms\ the effective excitation densities of these molecules are lower than their putative critical densities \citep[Table\,\ref{table_1} and ][]{Shirley2015}.  \ntwhp\ is a more reliable tracer of cold dense gas than CO, as CO freezes out more readily onto dust grains \citep{Caselli2022}. In addition to being a good tracer for high-density gas, \htwoco\ along with SO is a good tracer of shocked gas, while species such as \hcthn\ and \chthrcn\ with rotational transitions and $K$ ladder trace complex molecular chemistry typical of regions of high-mass star formation. Thus, a combination of observations of these molecules was chosen to characterise the large-scale filamentary molecular cloud as well as the high-mass star-forming condensation in the G183 region.

\subsection{Intensity maps of molecular line emission}
\begin{figure*}
    \centering
    \includegraphics[width=0.95\linewidth]{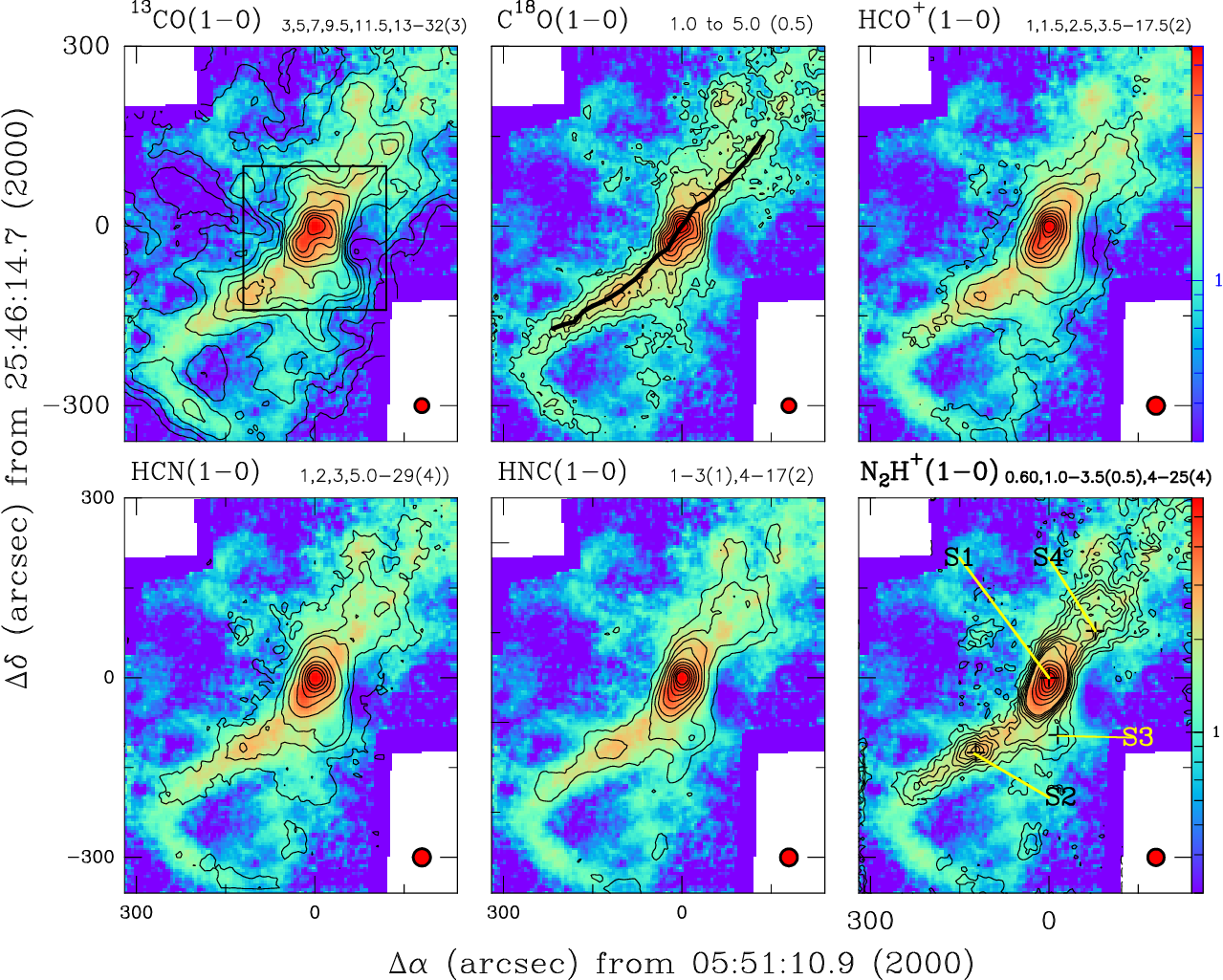}
    \caption{Integrated intensity maps in the 3\,mm band observed with the IRAM 30m telescope towards the G183 region. The colour plot in each panel is the integrated emission of \CeiO(1--0) and the contours correspond to molecular transitions, as is indicated above the panels
    in units of K\,\kms\ with step sizes in brackets. Above each panel the contour levels are given to the right.   The intensities are integrated between velocities of -12 to -2\,\kms\ , except for \ntwhp (1--0), which is integrated between -22 to 0\,\kms. In the top left panel, the region mapped at 1.4\,mm is shown by the smaller box.  Positions selected for further analysis are shown as '+' in the bottom right panel and marked as S1 (0\arcsec, 0\arcsec), S2(123\arcsec, -125\arcsec) and S3 (-15\arcsec,-96\arcsec).  The solid curve drawn on the \CeiO(1--0) map shows the direction in which the velocity analysis including the position-velocity diagram (Fig.\,\ref{fig_pvdiags}) was done. The HPBW is shown on the lower right.
    \label{fig_g183_90GHz_1}}
\end{figure*}
\begin{figure}
    \centering
    \includegraphics[width=0.95\linewidth]{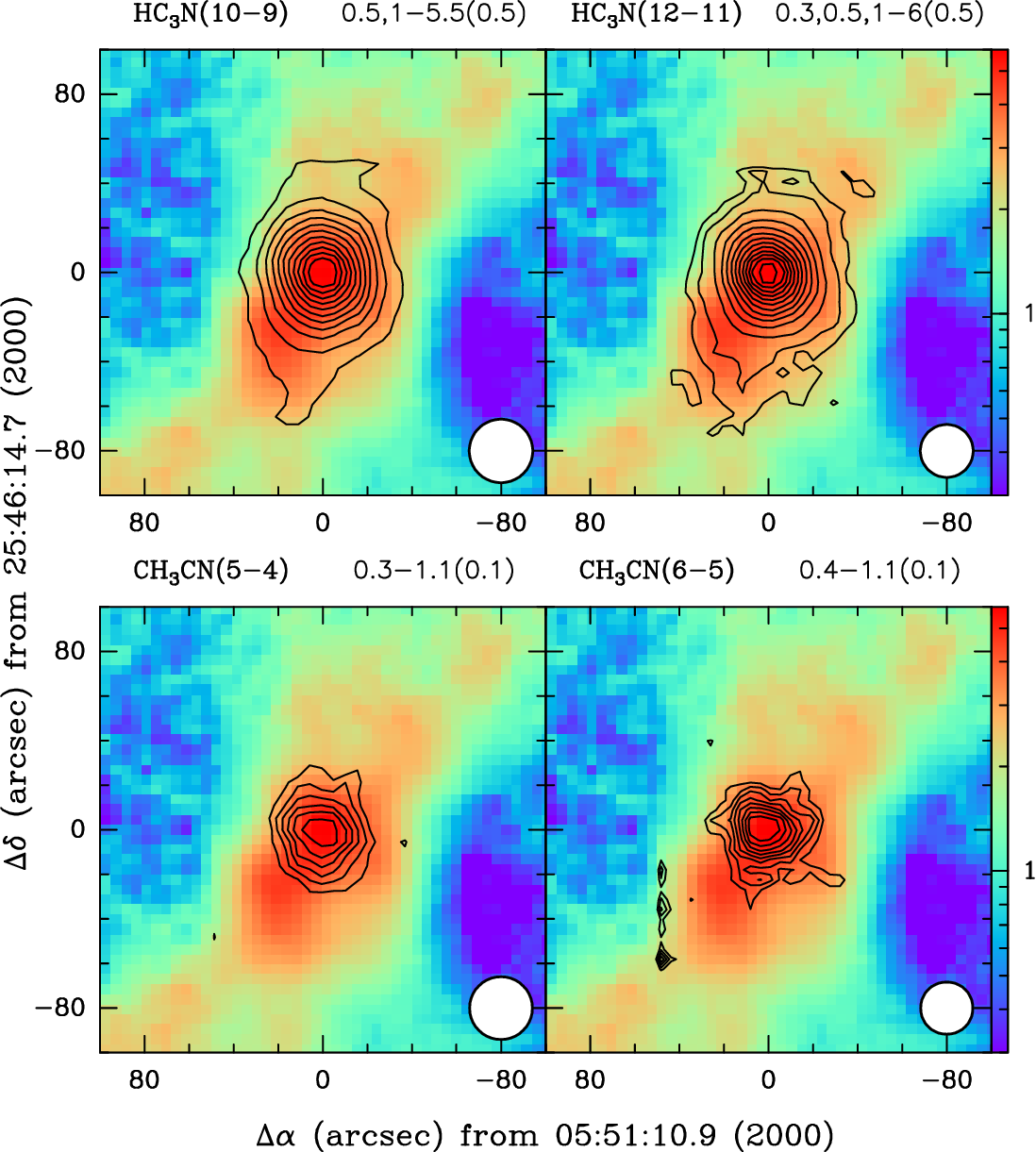}
    \caption{Integrated intensity maps of HC$_3$N and CH$_3$CN in the 3\,mm band observed with the IRAM 30m telescope.  Observations covered the region shown in Fig.\,\ref{fig_g183_90GHz_1} but emission in these lines  was only detected in the inner region shown here. The colour plot in each panel is the integrated emission of \CeiO(1--0) and contours correspond to molecular transitions as indicated above the panels. The intensities are integrated between velocities of -20 to 0\,\kms. Contour levels in units of (K\,\kms) are indicated at the top of the panels. The HPBW is shown on the lower right.
    \label{fig_g183_90GHz_2}}
\end{figure}
\begin{figure*}
    \centering
    \includegraphics[width=0.95\linewidth]{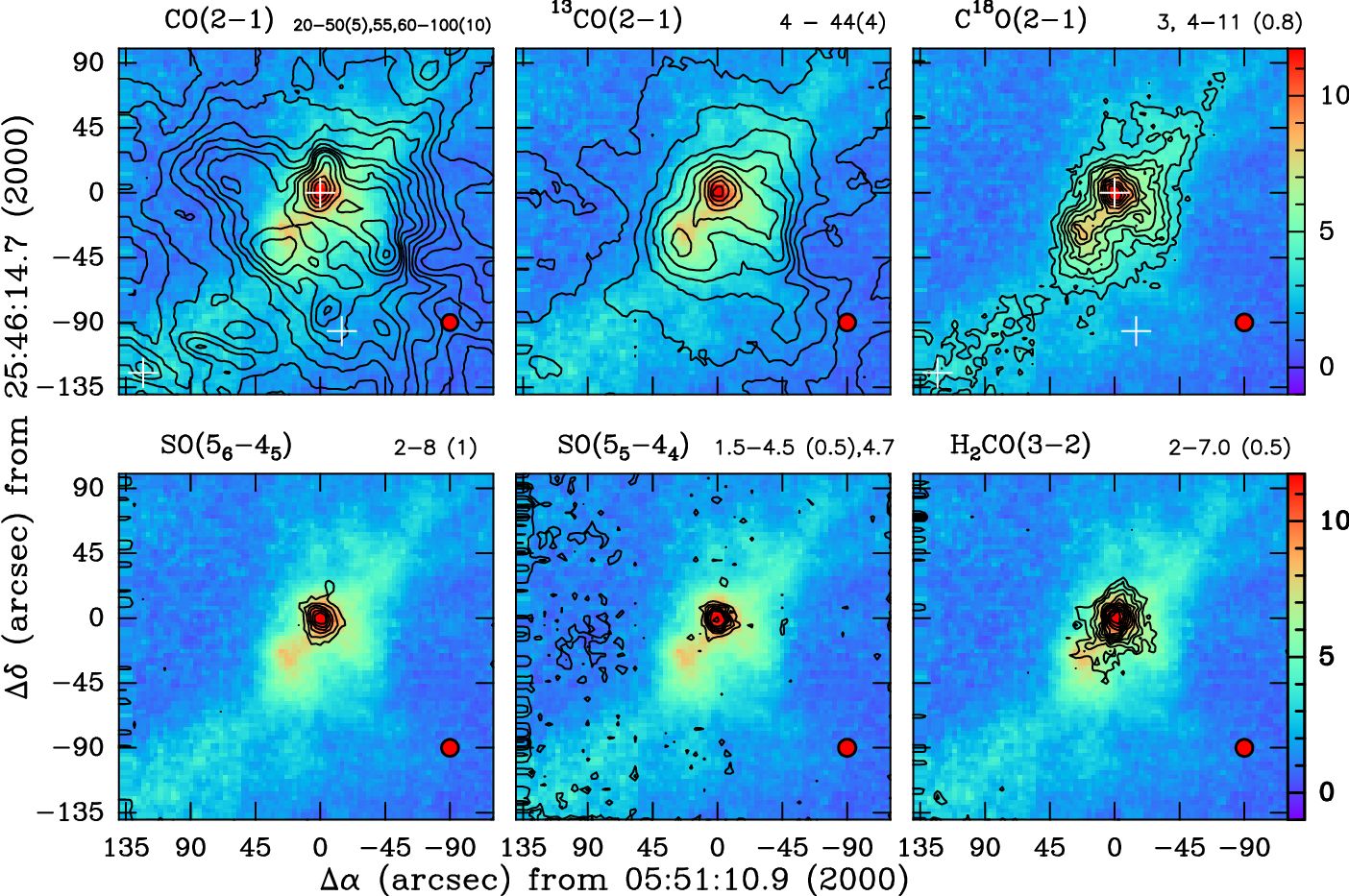}
    \caption{Integrated intensity maps at 1.4\,mm observed with the IRAM 30m telescope towards the G183 region. The colour plot in each panel is the integrated emission of \CeiO(2--1) and the contours correspond to molecular transitions, as is indicated above the panels. The intensities are integrated between velocities of -20 to 0\,\kms.  The HPBW is shown on the lower right.
    \label{fig_g183_220ghz}}
\end{figure*}

The 3\,mm molecular line emission from the G183 region shows that the molecular gas is distributed as a ridge or a filament, with a central condensation (source S1 hereafter) and multiple smaller clumps similar to the 250\,\micron\ continuum image (Fig.\,\ref{fig_g183_90GHz_1}). Of these molecular lines, the \thCO(1--0), \CeiO(1--0), and \ntwhp(1--0) emission trace, respectively, the long filament with increasing sensitivity to colder gas closer to the spine of the filament. The \hcop, HCN, and HNC transitions with higher critical densities detect the central condensation better, but are also detected from significant parts of the rest of the filament. The \ntwhp(1--0) clearly detects a second peak (source S2 hereafter) at an offset of (-123, 125), which coincides with a 70 and 160\,\micron\ source indicative of ongoing star formation.  The transitions due to the complex molecules such as \hcthn\ and \chthrcn\ with critical densities exceeding 10$^6$\,\cmcub\ are detected only from the central condensation (Fig.\,\ref{fig_g183_90GHz_2}) approximately 30\arcsec\ (0.3\,pc) in size. Figure\,\ref{fig_g183_220ghz} shows the integrated intensity maps of the spectral lines detected at 220\,GHz. The \CeiO(2--1) emission shows the central peak S\,1 detected at 3\,mm to be composed of two peaks: one at the exact location of S\,1 and the other to the south-east at (-21\arcsec, 27\arcsec). The SO and \htwoco\ emission are detected only within a region approximately 30--45\arcsec\ in radius around S\,1. The CO and \thCO\ (2--1) emission at 1.4\,mm is more extended towards the west than the emission of the 3\,mm CO lines. The east-west extension of emission is also seen in CO(3--2) but not in the emission of any of the other lines that we observed.

\subsection{Position-velocity diagrams}
\begin{figure}
    \centering
    \includegraphics[width=0.98\linewidth]{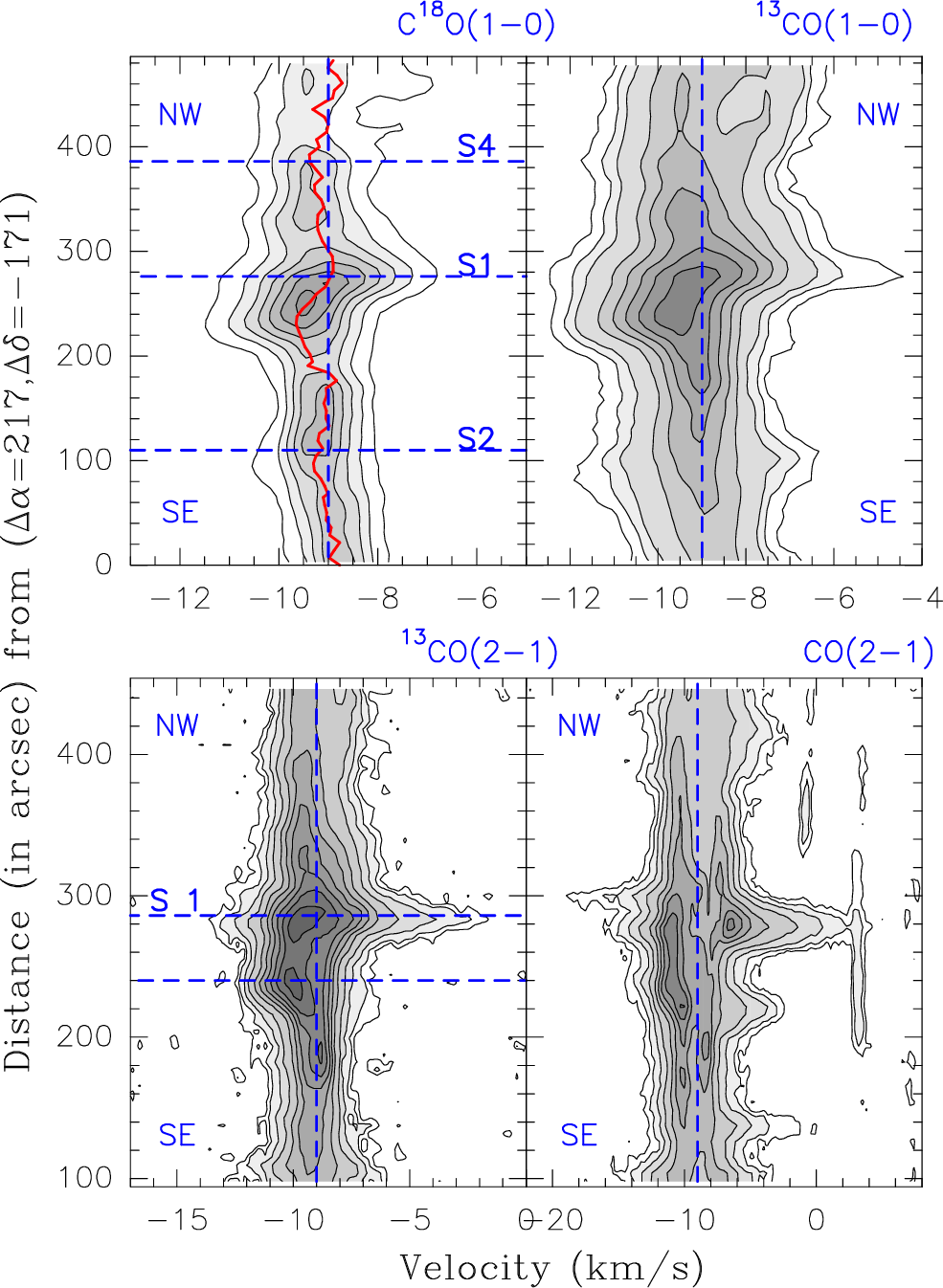}
    \caption{Position-velocity diagram for \CeiO, \thCO(1--0) and CO, \thCO(2--1) emission along the cut shown in Fig.\,\ref{fig_g183_90GHz_1}. The dashed blue lines are drawn at -9\,\kms\ to guide the eye.
    \label{fig_pvdiags}}
\end{figure}

The position-velocity ($p-v$) diagrams (Fig.\,\ref{fig_pvdiags}) derived for a cut along the filament (see Fig.\,\ref{fig_g183_90GHz_1}) show the emission from the south-eastern part of the filament to be centred at -9\,\kms, while the northern part is blueshifted by $\sim 1$\,\kms. In the star-forming hub, the outflowing gas is identified by the broad ($\Delta \upsilon \sim 20$\,\kms) lines particularly seen in the CO(2--1) emission, although the emission from S1 is heavily affected by optical depth effects. The \CeiO(1--0)  $p-v$ diagram captures the velocity along the spine of the filament, while the \thCO(1--0) shows broader emission since it captures some of the diffuse emission as well. The \thCO(2--1) $p-v$ diagram at a higher resolution identifies additional peaks along the filament. 

\begin{figure}[h]
    \centering
    \includegraphics[width=1.0\linewidth]{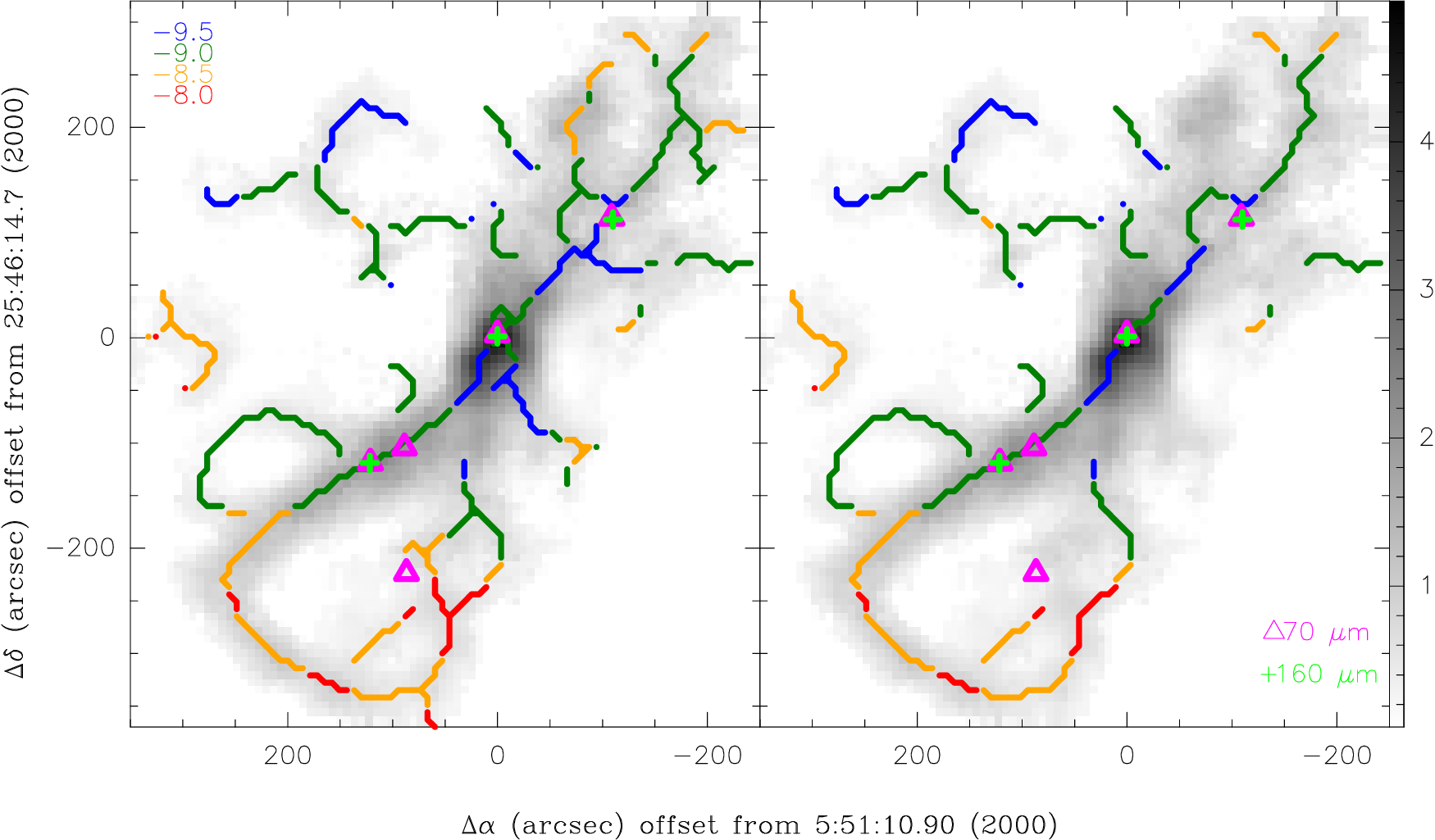}
    \caption{Velocity-coherent skeletons (left) and spines (right) identified using the automated tool {\sc CRISPY} applied to the \CeiO(1--0) datacube. A skeleton is defined as a ridge that has been gridded onto an image and a spine as a skeleton with all its branches removed. The greyscale image in each panel corresponds to the intensity map of \CeiO(1--0) integrated over -11 to -4\,\kms\ with S/N ratio $>5$. The continuously changing velocity centroids along the filaments are marked in different colours and the values are mentioned in the left panels. The triangle and '+' mark the positions of the 70\,\micron\ and 160\,\micron\ continuum sources, respectively.
    \label{g183_crispy_fil_SbyN_5}}
\end{figure}

In order to identify filaments in the p-p-v space, we used a Python-based automated  library CRISPY\footnote{\url{https://github.com/mcyc/crispy}} that detects density ridges in multidimensional data using the subspace-constrained mean shift (SCMS) algorithm \citep{chen2020} on the \CeiO(1--0) datacube. The SCMS algorithm finds ridges by moving walkers iteratively up the density field using a gradient ascent method. A ridge is defined as a smooth, continuous, one-dimensional object in a multi-dimensional density field. A skeleton is defined as a ridge that has been gridded onto an image and a spine as a skeleton with all its branches removed. For this work we have adopted a density threshold of 0.2\,K (five times the rms of the map) and a smoothing length of 1.5 pixels. Applying {\sc CRISPY} on voxels with signal-to-noise ratio $>5$, we identify the main filament along with multiple branches in the skeleton, with the southern part of the main filament being redshifted relative to the northern part (Fig.\,\ref{g183_crispy_fil_SbyN_5} ($left$)). Of these one branch extended approximately south-west to north-east joins the main filament exactly at the location of S1; however, it is too faint and is eventually discarded when the spine is identified by removing the bad branches  (Fig.\,\ref{g183_crispy_fil_SbyN_5} ($right$)). This establishes that the main filament is indeed a continuous structure in the p-p-v space, and hence is velocity-coherent.

\subsection{Analysis of spectra}

\begin{figure*}
    \centering
    \includegraphics[width=0.8\textwidth]{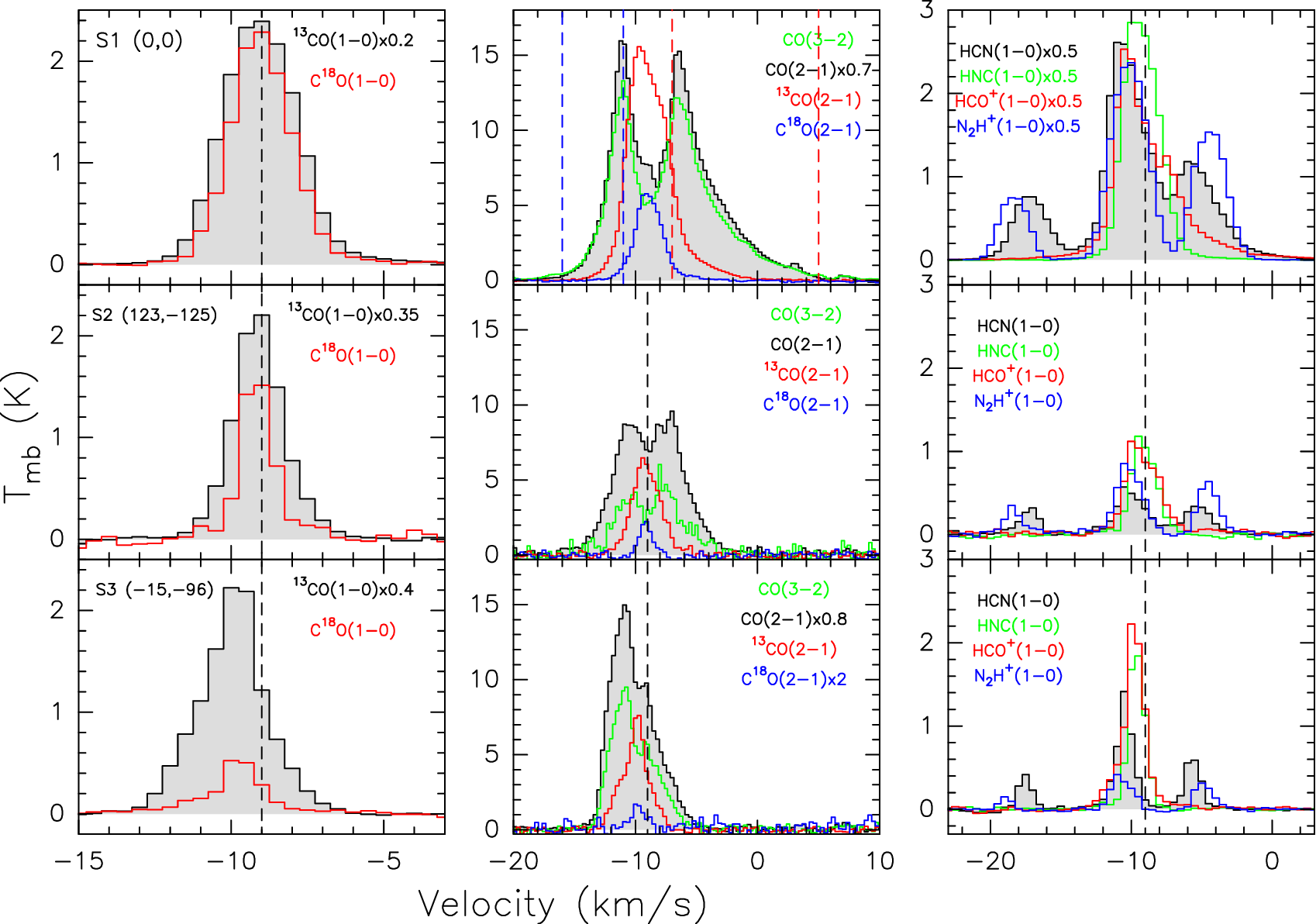}
    \caption{Comparison of spectra at selected positions (Fig.\,\ref{fig_g183_90GHz_1}) in G183. Panels in top, middle, and bottom rows correspond to positions S1, S2, and S3 with offsets (in arcseconds) (0,0), (-15,-96), and (123,-125) respectively. All 3\,mm spectra are convolved to a common resolution of 28\arcsec, while the 1.4\,mm spectra are convolved to a resolution of 14\,\arcsec. CO(3--2) has a native resolution of 15\arcsec. The vertical dashed line in the top row of the middle column panel shows the velocity ranges chosen for analysis of the blue and red wings of the spectra.}
    \label{fig_allspec}
\end{figure*}

\begin{figure}
    \centering
    \includegraphics[width=0.45\textwidth]{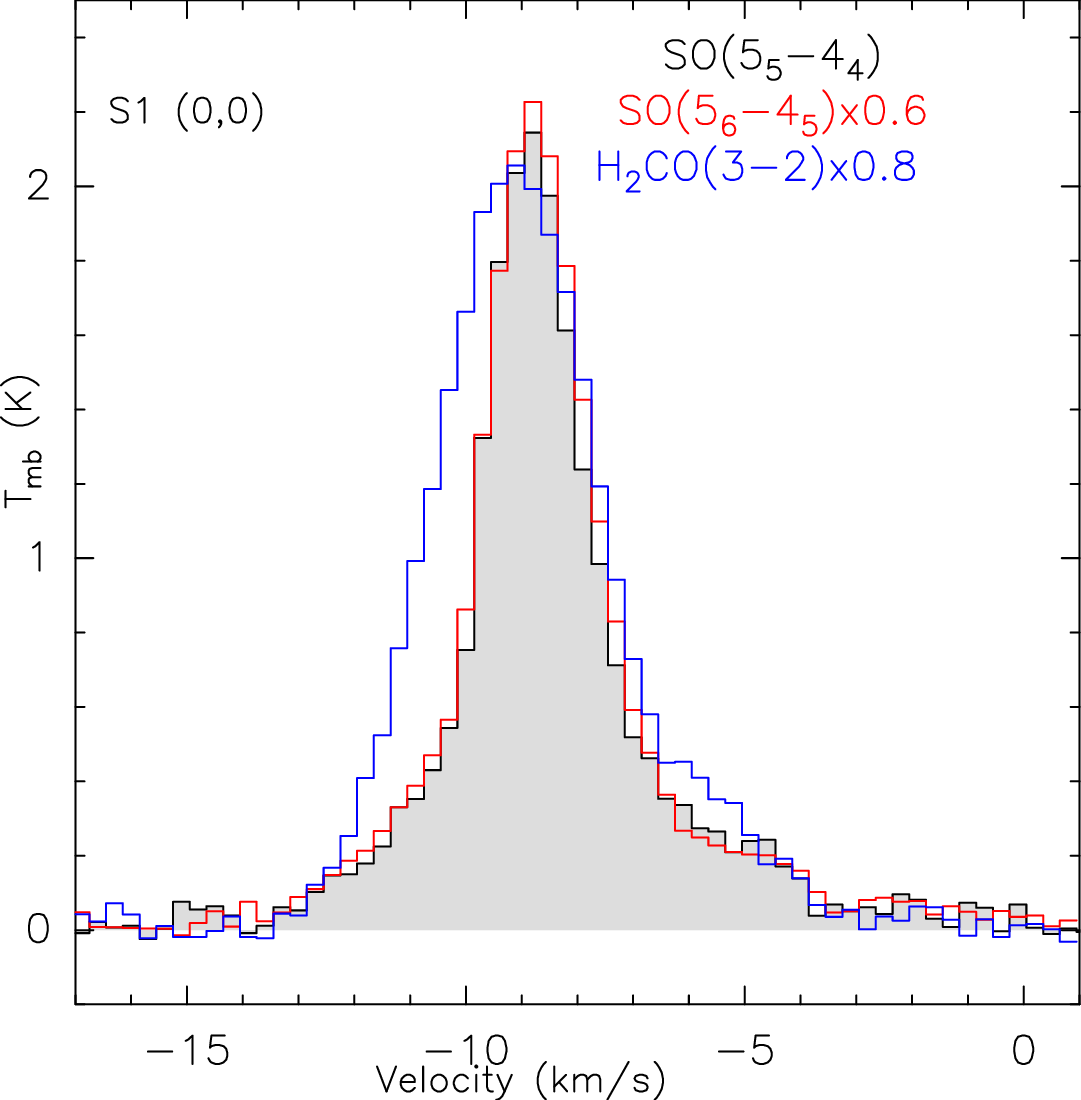}
    \caption{Comparison of spectra of the outflow tracers \htwoco(3--2), SO(5$_6$--$4_5$), and SO(5$_5$--$4_4$) at S1.}
    \label{fig_outfspec}
\end{figure}

We selected three positions in the region G183 for a comparison of the 1.4 and 3\,mm spectra of the different transitions at the same angular resolutions of 28\arcsec\ (3\,mm) and 14\arcsec (1.4\,mm) (Fig.\,\ref{fig_allspec}). The three positions  (Fig.\,\ref{fig_g183_90GHz_1}) correspond to the position of the source S1, the source S2, lying on the southern part of the filament also harbouring a far-infrared source and a peak lying to the south-west of S1 denoted by S3. The CO(2--1) and (3--2) spectra are optically thick, showing self-absorption, the \thCO\ lines show moderate optical thickness close to the centre of the line and the \CeiO\ lines are throughout single-peaked. The velocity of the absorption dip in the CO lines match exactly with the position of the \CeiO\ peak confirming that the dips are caused by opacity effects. At S1 CO(3--2), CO(2--1), \hcop(1--0) and HCN(1--0) lines are broadened due to the outflow (Fig\,\ref{fig_allspec}). Additionally, at S1 both \htwoco(3--2) and SO(5--4) spectra show strong outflow features with \htwoco(3--2) showing an enhanced blue tail, while the SO(5--4) profiles are more symmetric and narrower (Fig.\,\ref{fig_outfspec}). Position S3 (-15\arcsec,-96\arcsec) is not directly located on the filament but we chose it since it shows an enhanced blue emission in CO(2--1) and (3--2). This blue emission likely arises from the outflow due to the source S1 but as is seen in the channel map (Fig.\,\ref{fig_co21chanmap}) a significant fraction of it is due to an additional gas component centred around -12\,\kms. The emission feature at -12\,\kms\ is extended towards the south-west of the component approximately centred at -9\,\kms. The source S2, associated with a far-infrared continuum source, is best detected in the \ntwhp(1--0) map, and the CO(2--1) and (3--2) spectra are broad in velocity, clearly indicating the possibility of the presence of an outflow activity.   At all positions,  the HNC and \hcop(1--0) lines are centred at velocities same as the lines due to CO (and its isotopologues), while the \ntwhp(1--0) and HCN(1-0) lines are slightly blueshifted, peaking around -10.5 to -11\,\kms. A quantitative analysis of the spectra at the position of S1 to derive the physical parameters of the source is presented later in the paper (Sec.\,\ref{sec_cassis}).

\section{Interpretation of results}

\subsection{Core and filament mass}

Based on the detailed comparison of the spectra of CO and its isotopes, we conclude that the \CeiO\ emission is optically thin. We thus used the observed \CeiO(1--0) intensities to estimate the total mass in the filament as well as in central star-forming hub with S1. Assuming local thermodynamic equilibrium (LTE) and an excitation temperature of 20\,K,  we estimated the column density of the G183 region from the intensity of \CeiO(1--0) emission integrated between -15 to 0\,\kms\ using the following equation:

\begin{equation}
N({\rm C^{18}O}) = \frac{3h}{8\pi^3 \mu^2}\frac{Z}{\left[J_\nu(T_{ex})-J_\nu(T_{bg})\right]\left(1-\exp{(-{\frac{h\nu}{kT_{ex}}}})\right)}\displaystyle\int {T_{mb}dv}
,\end{equation}

with the $Z$ is the partition function given by
\begin{equation}
Z = \frac{k T_{ex}}{J_u B h}\exp\left(\frac{B J_u(J_u+1)h}{kT_{ex}}\right)  
,\end{equation}

where $J = \dfrac{h\nu/k}{\exp({h\nu/kT})-1}$, $B$ = 5.4891420$\times 10^{10}$\,s$^{-1}$ is the rotational constant for \CeiO, $\mu$ = 0.11079 D is its dipole moment, and $J_u$ is the upper level of the transition, equal to 1. Here, an excitation temperature of 20\,K is assumed for the gas. This is in most parts of the map slightly higher than the dust temperatures of the cold dust component, which are 10--20\,K (Sec.3). However, the effect on the derived gas column densities is minor (see below). Assuming a beam filling factor of unity and $T_{\rm ex}$ = 20\,K, we estimated the optical depth of the \CeiO(1--0) spectra using the following equation \citep{Shimajiri2014}:  

\begin{equation}
\tau({\rm C^{18}O}) = -\ln\left(1-\dfrac{T_{\rm MB}({\rm C^{18}O})}{5.27\left[J(T_{\rm ex}) - 0.1666\right]}\right)
,\end{equation}

where $T_{\rm MB}({\rm C^{18}O})$ corresponds to the value at the peak of the spectra at each pixel of the \CeiO(1--0) datacube. The values of $\tau$(\CeiO) lie between 0.002--0.02 and imply that our assumption of \CeiO(1--0) being optically thin is valid. For $T_{\rm ex}$ = 20\,K, the factor that converts the integrated \CeiO(1--0) intensities to $N$(\CeiO) is $1.7\times 10^{15}$ \cmsq\,K$^{-1}$\,[\kms]$^{-1}$ following Equation (1). A lower gas excitation temperature of 10\,K reduces the conversion factor and the derived column densities by only 12\%. We estimated $N$(H$_2$) from $N$(\CeiO) by adopting the $^{16}$O/$^{18}$O = 650 appropriate for a galactocentric distance of 10.44\,kpc \citep{WilsonRood1994} and CO/H$_2$=10$^{-4}$ \citep{Frerking1982}. The $N$(H$_2$) estimated from the \CeiO(1--0) intensity ranges between (1.3--6.3)$\times 10^{22}$\,\cmsq\ , with most of the filament away from the hub showing an almost uniform value between (1.95--2.6)$\times 10^{22}$\,\cmsq. This is lower than the value obtained from the grey-body fitting of dust continuum (Fig.\,\ref{fig_dcolden}) by almost a factor of 8--10 for an assumed gas-to-dust ratio of 220 \citep{Gianetti2017b}. A few possible sources of the discrepancy in the column density derived from gas and dust emission are (i) the dust continuum emission also being sensitive to a low-column-density diffuse warm dust component, unlike the \CeiO\ emission, and (ii) possible CO freeze-out onto dust grains when dust temperatures drop below $\sim 15$\,K \citep{Kramer1999,Caselli1999} along the filament, outside of the core region of massive star formation. The total mass of the part of the filament extended along the north-west--south-east direction is 2830\,\msun. The star-forming clump in the hub is better resolved in the 1.4\,mm maps so that it is possible to measure the spatial extent of the source more accurately. We thus used the \CeiO(2--1) intensities and Eq.(1) to  estimate the mass of the clump to be 156\,\msun. 

\subsection{Line intensity ratios}

Recent studies of many of the molecular lines traditionally used as tracers of high-density gas show that these molecular lines trace much lower densities than their critical densities \citep[e.g.,][]{Kauffmann2017, Pety2017}. In comparison to the critical density ($n_{\rm crit}$) calculated under the simplifying assumptions of two-level systems and optically thin emission, it is more realistic to use the effective excitation density ($n_{\rm eff}$) defined as the density at which a modest (1\,K\,\kms) line intensity is observed for a given gas kinetic temperature and column density \citep{Shirley2015}. Table\,\ref{table_1} also lists for all the spectral lines in our data $n_{\rm crit}$ and $n_{\rm eff}$ estimated assuming a kinetic temperature of 20\,K \citep{Shirley2015}.  Among the four dense gas tracers  (HCO$^+$, HCN, HNC, and N$_2$H$^+$) that we observed at 3\,mm, we see that HCO$^+$(1--0) has the smallest $n_{\rm eff}$, and also traces the most extended structure, while N$_2$H$^+$(1--0) has the highest $n_{\rm eff}$, and hence has the most compact emission. We also note that the emission from the \hcthn\ and \chthrcn\ transitions at 1.4\,mm with $n_{\rm eff}\sim 10^5$\,\cmcub\ are confined to the core around S1, indicating the presence of very high-density gas in the region. 

Integrated intensity ratios of molecular lines are often used to trace molecular gas properties, which can be linked to star formation activities and many surveys have been carried out in our Galaxy and towards nearby galaxies in the 3\,mm band \citep[e.g.][]{Wang2020,Jimenez2019}. In G183 we have estimated the integrated intensity ratios HCN(1--0)/HNC(1--0), HCN(1--0)/\hcop(1--0), \hcop(1--0)/\thCO(1--0), HCN(1--0)/\thCO(1--0), \ntwhp(1--0)/\thCO(1--0), and HNC(1--0)/\thCO(1--0) for all pixels with intensities $>1$\,K\,\kms. 

\begin{table}[h]
\caption{Ratios of integrated intensities of (1--0) transitions of species indicated. \label{tab_ratio}}
\centering
\begin{tabular}{lcrr}
\hline \hline
Ratio & Range & Mean & Galactic Sources$^a$\\
\hline
HCN/HNC & 0.8--1.80 & 1.1 & 1.95--2.78 \\
HCN/\hcop & 0.51--2.48 & 1.0 & 1.09--1.29\\
HCN/\thCO & 0.08--0.86 & 0.2 & 0.15--0.35\\
\hcop/\thCO & 0.08--0.57 & 0.2 & 0.15--0.37\\
HNC/\thCO & 0.06--0.51 & 0.2 & 0.08--0.13\\
\ntwhp/\thCO & 0.04--0.88 & 0.2 & 0.06-0.10\\
\hline
\end{tabular}
\tablefoot{$^a$ Values for GMF54 and W51 arm \citep{Wang2020}}
\end{table}

The intensity ratios measured for G183 compare well with the values in Galactic sources observed at a similar spatial resolution as the current observations (Table\,\ref{tab_ratio}). We have used \thCO(1--0) with the maximum emitting volume as representative of the total molecular gas content in the region. Among the ratios of transitions considered, based on the current understanding significant emission of the HCN(1--0), HNC(1--0) and \hcop(1--0) could arise from gas with densities $\sim 10^3$\,\cmcub, and \ntwhp(1--0) at 10\,K with an effective excitation density of $10^4$\,\cmcub\ likely traces the dense gas in the filament the best. The HCN/HNC ratio has been used as an indicator of evolutionary stages of the molecular cloud, since at temperatures above 30\,K HNC starts to convert more efficiently to HCN \citep{hacar_2020}. In G183 the HCN/HNC ratio does not show much variation in the region with highest (1.8) being around S1. Based on the empirical relation between kinetic temperature and the HCN/HNC ratio assuming both lines to be optically thin \citep{hacar_2020}, the observed HCN/HNC would indicate a temperature of $<20$\,K throughout G183. However it is clear that owing to the presence of S1 the temperature in the region immediately next to it must be higher. A possible reason for the observed low HCN/HNC ratios could lie in the HCN(1--0) line being optically thick. In the absence of data of a rarer isotope of HCN or high-$J$ transitions with higher critical densities, it is not possible to resolve this discrepancy.

\subsection{Kinematic properties of the main filament}

\begin{figure}[h]
    \centering
    \includegraphics[width=0.5\textwidth]{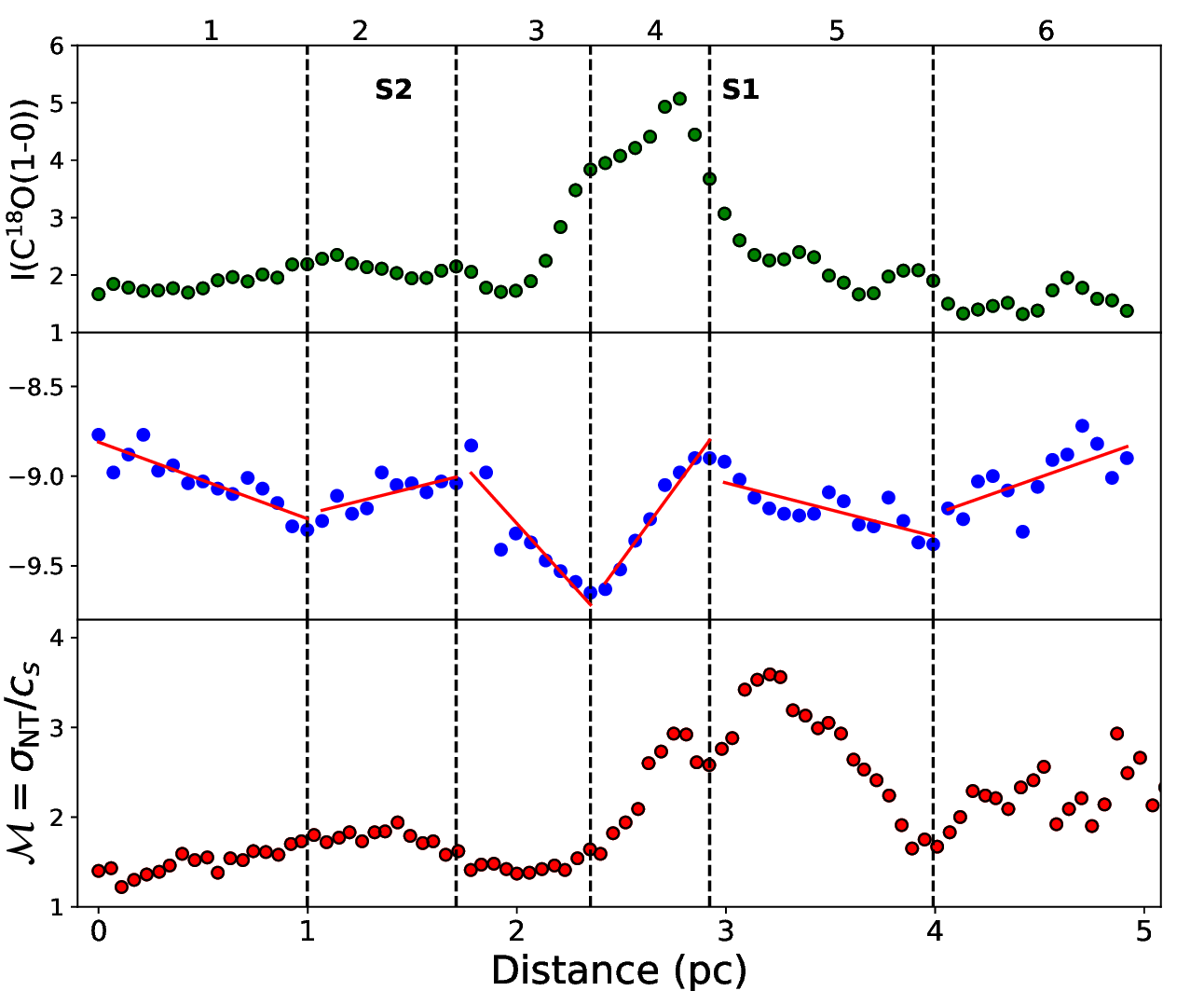}
    \caption{Variation in intensity ($I$(\CeiO(1--0)) in K\,\kms, intensity-weighted velocity centroid ($v_{\rm centroid}$) in \kms\ , and Mach number ($\mathcal M = \sigma_{\rm NT}/c_s$ ) along the length of the filament shown in Fig\,\ref{fig_g183_90GHz_1}. $\sigma_{\rm NT}$ is the non-thermal velocity dispersion (Eq. 3) and $c_s$ is the sound speed estimated at 20\,K. The velocity centroids in multiple sections of the filament are fitted with straight lines to determine the gradients.}
    \label{fig_velofit}
\end{figure}

Figure\,\ref{fig_velofit} shows the variation in the integrated  line intensity and the intensity-weighted centroid velocity along the filament identified in the \CeiO(1--0) $p$-$v$ diagram (Fig.\,\ref{fig_pvdiags}). We analyzed the velocity profile by splitting the total length (5\,pc) of the filament into six parts. The mean centroid velocity (v$_{\rm centroid}^{\rm mean}$) in parts 1, 2, 5, and 6 are almost the same (-9.0\,\kms) with a dispersion of the centroid velocity ($\sigma$ (v$_{\rm centroid}$)) of 0.1-0.2\,\kms\ (Table\,\ref{tab_velofit}). For parts 3 and 4, v$_{\rm centroid}^{\rm mean}$ is -9.3\,\kms\ and $\sigma$ (v$_{\rm centroid}$) is 0.3\,\kms.  We estimated the velocity gradients along the length of the filament by fitting straight lines to the velocity profile part-wise. Parts 1, 2, 5, and 6 approximately show a velocity gradient between 0.3--0.5\,\kms\,pc$^{-1}$. Parts 3 and 4 lying approximately on either side of S1 show much stronger velocity gradients of -1.3 and 1.5\,\kms\,pc$^{-1}$ with a change of sign (Fig.\,\ref{fig_velofit}). 

We identify the velocity pattern in parts 1, 2, 5, and 6 as being dominated by the large-scale and quiescent velocity distribution, whereas kinematics of parts 3 and 4 is dominated by the accretion and outflow activities associated with S1. The V-shaped velocity profile along with a correlated intensity peak is consistent with the inflow of gas due to the strong gravitational potential of a high-mass YSO \citep{sen_2024}. Such V-shaped velocity profiles could also arise from either cloud-cloud collisions \citep{Fukui2018} or from filaments forming within large-scale sheets compressed by propagating shock fronts \citep{Arzoumanian2018,Arzoumanian2021}. In the cloud-cloud collision scenario typically two distinct velocity components connected by a bridge spanning a few to 20\,\kms\ are observed \citep[e.g.,][]{Haworth2015,Liow2020,Fukui2021}. In G183, we do not see any evidence of a bridge-feature or broad shocked emission either in the spectra (at resolutions down to 5\arcsec\ NOEMA observations (Mookerjea et al. in prep) or in the p-v diagram. Comparison with the outcome of the sheet scenario is similarly inconclusive. We note that the observed velocity separation is subject to uncertainties due to projection effects. Within this caveat, the relatively quiescent kinematics and lack of multiple velocity components make gas inflow along the filament and accretion to the star-forming core the most plausible origin of the observed V-shaped structure in G183. 

We estimated the mass accretion rate of a filament segment using $\dot{M} = v_{\rm grad} M/\tan{\theta}$ \citep{kirk_2013}, where $M$ is the mass of the segment, $v_{\rm grad}$ is the velocity gradient across it and $\theta$, the angle made by the filament with the plane of the sky is assumed to have an average value of 45$^{\circ}$. For segments 3 and 4, respectively, $v_{\rm grad}$  are -1.3 and 1.5 \kms\,pc$^{-1}$, the masses are 173 and 240\,\msun\ (assuming 20\,K and LTE), and hence $\dot{M_{\rm acc}}$ -3.2$\times 10^{-4}$ and 5.4$\times 10^{-4}$\,\msun\,yr$^{-1}$. Thus we estimated a total mass accretion rate of 8.6$\times 10^{-4}$\,\msun\,yr$^{-1}$ for the source S1. The mass accretion rates typically found in low-mass protostars range from a few 10$^{-8}$ to a few times 10$^{-4}$\,\msun\,yr$^{-1}$ \citep{Fiorellino2023}, with the HFSs harbouring massive protostars showing values up to 1.5$\times 10^{-3}$\,\msun\,yr$^{-1}$ \citep{Hu2021,Ma2023}. The velocity gradient created by a core of mass ($M$) 156\,\msun\ within a radius ($R$) of 0.18\,pc (17\farcs7 at 2.1\,kpc), given by $\displaystyle\sqrt{GM/2R^3}$ is equal to 7.6\,\kms\,pc$^{-1}$. The velocity gradients obtained from the \CeiO(1--0) data are about one fifth of the velocity gradient that the core mass could generate. We note that the velocities measured are along the line of sight and the inclination of the filament is not known; hence, the actual values could be larger than the observed values. On the other hand, at a resolution of 0.1--0.2\,pc it is not possible to distinguish between the large-scale velocity fields from the accretion flows; hence, the velocity gradients due to accretion in reality could be smaller than the observed values.

\begin{table}
\caption{Analysis of velocity profile Figure \ref{fig_velofit} \label{tab_velofit}}
\centering
\begin{tabular}{rcrrr}
\hline \hline
Fil part & v$_{grad}$ & v$_{\rm centroid}^{\rm mean}$ & $\sigma$ (v$_{\rm centroid}$)& M/L\\
&(\kms\,pc$^{-1}$) &(\kms) & (\kms) &(\msun\,pc$^{-1}$)\\
\hline
1 & -0.43$\pm$0.06 & -9.0 & 0.13 & 194\\
2 & 0.42$\pm$0.09 & -9.1 & 0.10 & 312\\
3 & -1.28$\pm$0.21 & -9.3 & 0.25 & 727 \\
4 & 1.54$\pm$0.11 & -9.3 & 0.29 & 554\\
5 & -0.34$\pm$0.06 & -9.1 & 0.14 & 308\\
6 & 0.47$\pm$0.12 & -9.0 & 0.18 & 369\\
\hline
\end{tabular}
\end{table}

The total internal velocity dispersion of a cloud was determined by the combined contribution of the large-scale velocity variations and the small-scale thermal and non-thermal motions. The 5-pc long G183 filament provides an opportunity to study each of the components individually. We estimated the non-thermal velocity dispersion along the line of sight ($\sigma_{\rm NT}$) from the velocity-deconvolved full width half maximum (FWHM; $\Delta V$) by subtracting the thermal contribution to line broadening as

\begin{equation}
    \sigma_{\rm NT} = \displaystyle\sqrt{\dfrac{\Delta V^2}{8\ln 2} - \dfrac{kT_{\rm kin}}{m}}
,\end{equation}

where $\sigma_{\rm th} = \sqrt{kT_{\rm kin}/m}$ is the thermal velocity dispersion of a tracer with molecular weight $m$ and at a kinetic temperature of $T_{\rm kin}$. At 20\,K, the (isothermal) sound speed ($c_s = \sqrt{kT_{\rm kin}/\mu m_{\rm H}}$, with $\mu=2.33$, \citet{hacar_2023}) is 0.27\,\kms. 

Based on Gaussian fitting of individual \CeiO(1--0) spectra along the same cut, we estimated the FWHM ($\Delta V$) for all positions along the filament. We find that in segments 1--3 of the filament $\Delta V$ lies between 1.0--1.1\,\kms, in segments 4 and 5 it lies between 1.7--2.2\,\kms\  with an average of 1.9\,\kms\ and in segment 6 it is $\sim 1.4$\,\kms. These are significantly larger than the thermal broadening at 20\,K, and hence indicate non-thermal contributions possibly arising due to turbulence. Figure\,\ref{fig_velofit} also shows the variation in the observed Mach number, $\mathcal{M}$, defined as the ratio of non-thermal velocity dispersion ($\sigma_{\rm NT}$) with $c_s$ of \CeiO\ at 20\,K. The Mach number is classically used to distinguish between the sonic ($\mathcal{M}\leq 1$), transonic ($1 <\mathcal{M}\leq 2$), and supersonic ($\mathcal{M} > 2$) regimes in isothermal non-magnetic fluids. Starting from the south up to 2.4\,pc along the filament  $\mathcal{M}$ lies between 1.0--1.8, suggesting the flow in this part is transonic. Between 2.4 to 3.7\,pc, the flow is highly supersonic ($2<\mathcal{M}<3.5$) and further to the north it is mildly supersonic.

We explored the stability of segments of the filament by studying their line mass, $M/L$ (mass per unit length), vis a vis the critical line mass. Considering the thermal and turbulence pressure support the critical line mass that describes the stability of such a system is given by \citep{wang2014}

\begin{equation}
M_{\text{crit}}^{\text{line}} = \left[1+\left(\frac{\sigma_{\text{NT}}}{c_s}\right)^2\right] \left[16\, \text{M}_{\odot}\,\text{pc}^{-1}\times\left(\frac{\text{T}}{10\,\text{K}}\right)\right]
.\end{equation}

Based on the values of $\mathcal M$ calculated earlier, for a temperature of 20\,K we estimated the $M_{\rm crit}^{\rm line}$ for the different segments of the filament to be 135, 176, 208, and 367 \,\msun\,pc$^{-1}$ for $\mathcal{M}$ = 1.5, 1.8, 2, and 2.8, respectively. This implies that for most of the filament  segments, the average $M/L$ likely exceeds the estimated $M_{\rm crit}^{\rm line}$ considering only thermal and turbulent pressure support. However, we do not detect any star formation spurt as would be expected if the filament segments were indeed unstable against gravitational collapse. We envisage two possible scenarios: (a) the resolution of the \CeiO(1--0) map likely does not allow us to measure the gas very close to the spine of the filament, and hence it can not detect any enhanced turbulence generated by accretion, and (b) our analysis of the stability of the filament does not consider the role of magnetic pressure, which likely exists given that most Galactic ISM filaments are strongly threaded by magnetic field lines \citep{Pattle2023}. 

The equilibrium solution as above is prone to fragmentation due to gravity \citep[see e.g.][]{Inutsuka1997}. In the presence of non-thermal motions, the separation between clumps formed due to fragmentation is given by \citep{wang2014}
    \begin{equation}  
    \lambda_{\rm clump} = 1.24\, {\rm pc} \left[\ \frac{\sigma_{\rm tot}}{1\,{\rm km s^{-1}}}\right]\left[\ \frac{n_{c}}{10^5 {\rm cm^{-3}}}\right]^{-1/2}
    ,\end{equation}

where $\sigma_{\rm tot} = \dfrac{\Delta V}{\sqrt{8\ln 2}}$ and $n_{\rm c}$ is the number density at the centre of the filament. 

 The measured values of $\Delta V$ = 1.2, 2.0, and 1.5\,\kms\ in the three identified velocity regimes of the filament correspond to $\sigma_{\rm tot}$ = 0.5, 0.8, and 0.6\,\kms\ , respectively. For a cloud with $n_c = 10^4$\,\cmcub, the clump distances would thus be 1.9, 3.0, and 2.3 pc. Although the $M/L$ values estimated above suggest the need for magnetic support, we find that the distance between the sources S1 and S2 is 1.8\,pc, which compares well with the values of fragmentation length scale dictated by thermal and turbulent support.

\subsection{Velocity structure function}

\begin{figure}[h]
    \centering
    \includegraphics[width=0.50\textwidth]{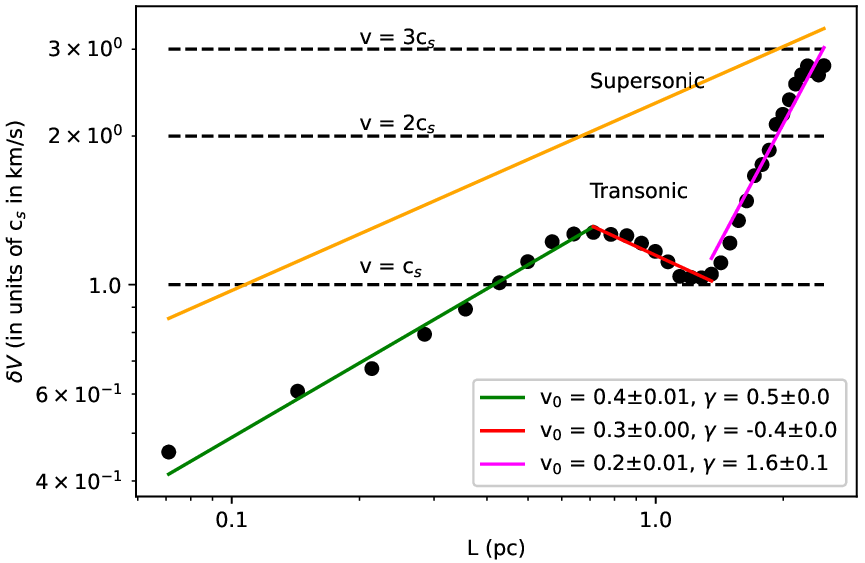}
    \caption{Plot of average velocity dispersion, $\delta V = <|v(r)-v(r+L)|^{2}>^{1/2} = v_{0}L^{\gamma}$, in units of sound speed $c_s$ at 20 K (0.27 \kms\,) along the main filament starting from the south. The straight yellow line corresponds to the Larson's law with $v_0 = 0.63$ and $\gamma = 0.38$ \citep{Larson1981}. }
    \label{fig_vstructfunc}
\end{figure}

In order to quantify the contribution of velocity changes and large-scale velocity gradients to the macroscopic motions inside the filament, we analysed the variation in line ncentroids using the velocity structure function. This analysis is similar to the analysis performed by \citet{Hacar2016} on the 6-pc-long Musca filament. The n$^{th}$ order velocity structure function is defined as $S_n(L) = <|v(r)-v(r+L)|^n>$ and the square root of the second-order structure function gives an average velocity dispersion along the filament with respect to the filament length.
\begin{equation}
\delta V=[S_{2}(L)]^{1/2} = <|v(r)-v(r+L)|^{2}>^{1/2} = v_0L^\gamma
,\end{equation}

where $v(r)$ is the velocity at a position on the filament, and  $v_0$ and $\gamma$ are the scaling coefficient and power-law index, respectively. 
We estimated the second-order structure function in velocity as a function of length $L$ (i.e. lag) from the \CeiO(1--0) observations along the main axis of the G183 filament (Fig.\,\ref{fig_vstructfunc}).  We used the intensity-weighted centroid velocities (Fig.\,\ref{fig_velofit}) for this purpose. A single linear fit to the second-order structure function across all length scales leads to $\gamma = 0.4$; v$_0$ = 0.4\,\kms, a power-law index matching well with Larson’s velocity dispersion-size relationship, $\delta V$ = 0.63 $L^{0.3}$ \citep{Larson1981}. However, the variation in the second-order structure function for the filament appears to be more consistent with a broken power law. A linear fit at correlation lags $L\leq 1$\,pc within the G183 filament produces a (tran-)sonic structure function with a power law index of $\gamma = 0.5 $ and v$_0$ = 0.4\,\kms. Beyond $L > 1.2$\,pc the structure function shows a steep power-law index of $\gamma = 1.8$ and v$_0$ = 0.2\,\kms\ and  becomes supersonic ($\delta V > 2 c_s$) for $L>2$\,pc. The rapid increase in the slope of $\delta V$ at characteristic scales of $\sim 1.2$\,pc suggests a change in the internal velocity field of the G183  filament and can be explained in terms of the ordered velocity gradients created by the accreting source S1 towards the north of the filament. The velocity dispersion in the filament on smaller scales are (tran-)sonic with $\gamma=0.5$, which is also consistent with Larson's relation, although the v$_0$ is significantly less than the predictions of Larson's law. At lags between 0.7 to 1.2\,pc, we obtain a negative power-law index with a magnitude consistent with Larson's relation. 

\subsection{Outflow activity from S1}

\begin{figure}[h]
    \centering
    \includegraphics[width=0.49\textwidth]{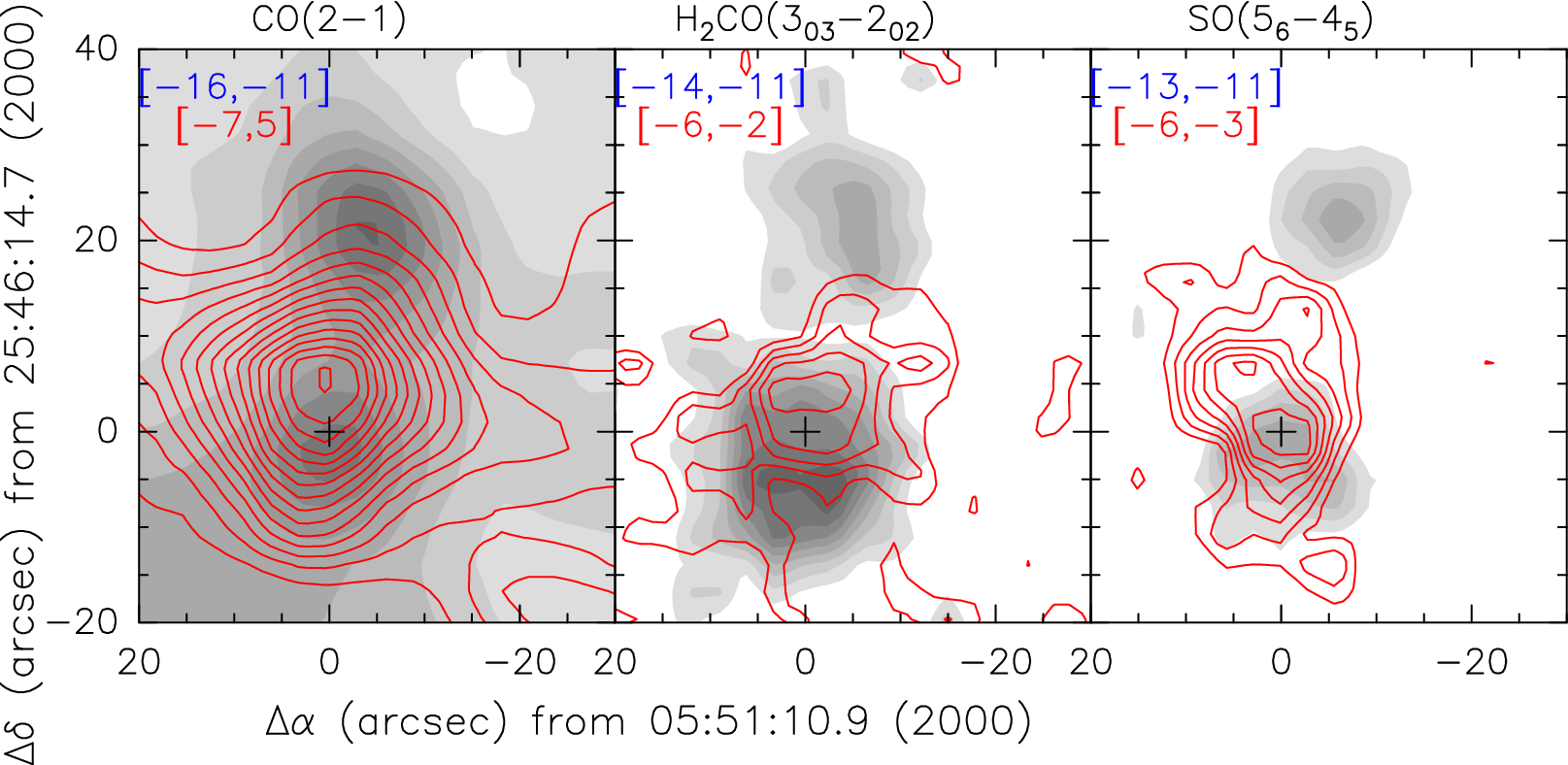}
    \caption{Distribution of the blue (greyscale) and red lobes (contours) of outflowing gas as detected in some of the transitions. The '+' marks the position of the massive YSO. Contour levels are in K\,\kms\ . For CO(2--1) blue contours are at 10 to 45 in steps of 5 and red contours are at 15 to 85 in steps of 5. For \htwoco(3--2) blue contours are at 0.3 to 1.0 in steps of 0.1 and  red contours are at   0.3 to 0.7 in steps of 0.1. For  SO($5_6$--$5_4$) blue contours are at 0.3 to 0.1 in steps of 0.1 and red contours are at 0.3 to 1.4 in steps of 0.1. In all panels the  velocity ranges used for integration for the blue and red lobes are marked.
    \label{fig_outfmap}}
\end{figure}

The detection of broad emission spectra consistent with outflow activity confirm the presence of the embedded massive protostellar object S1 in the most massive clump located on the main filament in G183. The distribution of the redshifted and blueshifted parts of the outflow as detected using CO(2--1), \htwoco(3--2) and SO($5_6$--$4_5$) reveal a red and a blue lobe approximately overlapping with each other at S1 (Fig.\,\ref{fig_outfmap}). This is suggestive of an outflow that is almost completely aligned with the line of sight of the observer. We notice that the blue lobe is significantly weaker than the red lobe at S1, but has a much brighter counterpart further up north that is detected in all the tracers.  Considering the fact that there is no additional continuum source that can support the northern blue lobe and that the two blue lobes are connected by an outer common envelope of contours in CO(2--1), which is the most easily excited, it is likely that these two lobes are parts of the same outflow. A possible explanation for the significant difference in the appearance of the two lobes of the outflow  could lie in the difference in the density of molecular gas towards and away from the observer. The red lobe of the outflow appears to be quite compact and more entrained, possibly because it is advancing in a denser medium than its blue counterpart. The ambient medium in which the blue lobe could be advancing towards the observer appears to be inhomogeneous as well as of lower density. Close to S1 the blue lobe has possibly cleared significant material and advanced to the part with the larger bulk of emission that is further north than the red lobe (relative to S1). Based on the appearance of the red lobe, we consider the inclination of the outflow to be almost aligned with our line of sight. We calculated an excitation temperature of 30\,K at LTE for S1 assuming optically thick CO(3--2) and CO(2--1) emission at the centre of the line.  Comparing the \twCO(2--1) and \thCO(2--1) intensities in the red (-7 to 5\,\kms) and blue (-16 to -11\,\kms) wings, we estimated the optical depth ($\tau_{12}$) of the \twCO(2--1) using the formula
\[
\dfrac{T_{\rm mb,12}}{T_{\rm mb,13}} = \dfrac{1-\exp(-\tau_{12})}{1-\exp(-\tau_{12}/R)}
,\]
where $R$ is the $^{12}$C/$^{13}$C abundance ratio of 85.9 for the  galactocentric distance of 10.44\,kpc \citep{WilsonRood1994} of G183. The optical depth $\tau_{12}$ was estimated at each velocity channel and a mean value of 2 and 3 was determined for the blue and red wings of the CO(2--1) emission. The opacity-corrected CO(2--1) main beam temperature in the wings estimated as $T_{\rm mb}^{\rm corr}$ = $T_{\rm mb}\tau_{12}/\left(1-\exp{\left(-\tau_{12}\right)}\right)$ was used to derive the column density and mass of the molecular gas in the outflow. 

In the absence of any information on the inclination of the outflow, we did not correct for the inclination while evaluating various parameters of the outflow. Appendix\,\ref{sec_outpar} outlines the details of the definitions of the outflow parameters calculated here. We calculated the masses of the blue ($M_b$) and red ($M_r$) lobes to be 1.9 and 0.96\,\msun\ , respectively, leading to a total outflow mass ($M_{\rm out}$) of 2.86\,\msun. The characteristic timescale ($t$) of the outflow was estimated to be equal to 1.1$\times 10^4$\,yr. The mass entrainment rate of the molecular outflow is equal to 2.6$\times 10^{-4}$\,\msun\,yr$^{-1}$. The total momentum of the outflow is 33\,\msun\,\kms\ , while energy of the outflow is equal to 4.1$\times 10^{45}$\,erg. The mechanical force ($F_{\rm m}$) and mechanical luminosity ($L_{\rm m}$) were calculated to be equal to 3$\times 10^{-3}$\,\msun\,\kms\,yr$^{-1}$ and 3.1\,\lsun\ , respectively. The outflow parameters thus derived are within the range of values seen in high-mass star-forming regions in the inner Galaxy \citep{Beuther2002}.

\section{LTE modelling of the molecular gas in S1 \label{sec_cassis}}

\begin{figure*}
    \centering
   \includegraphics[width=0.90\linewidth]{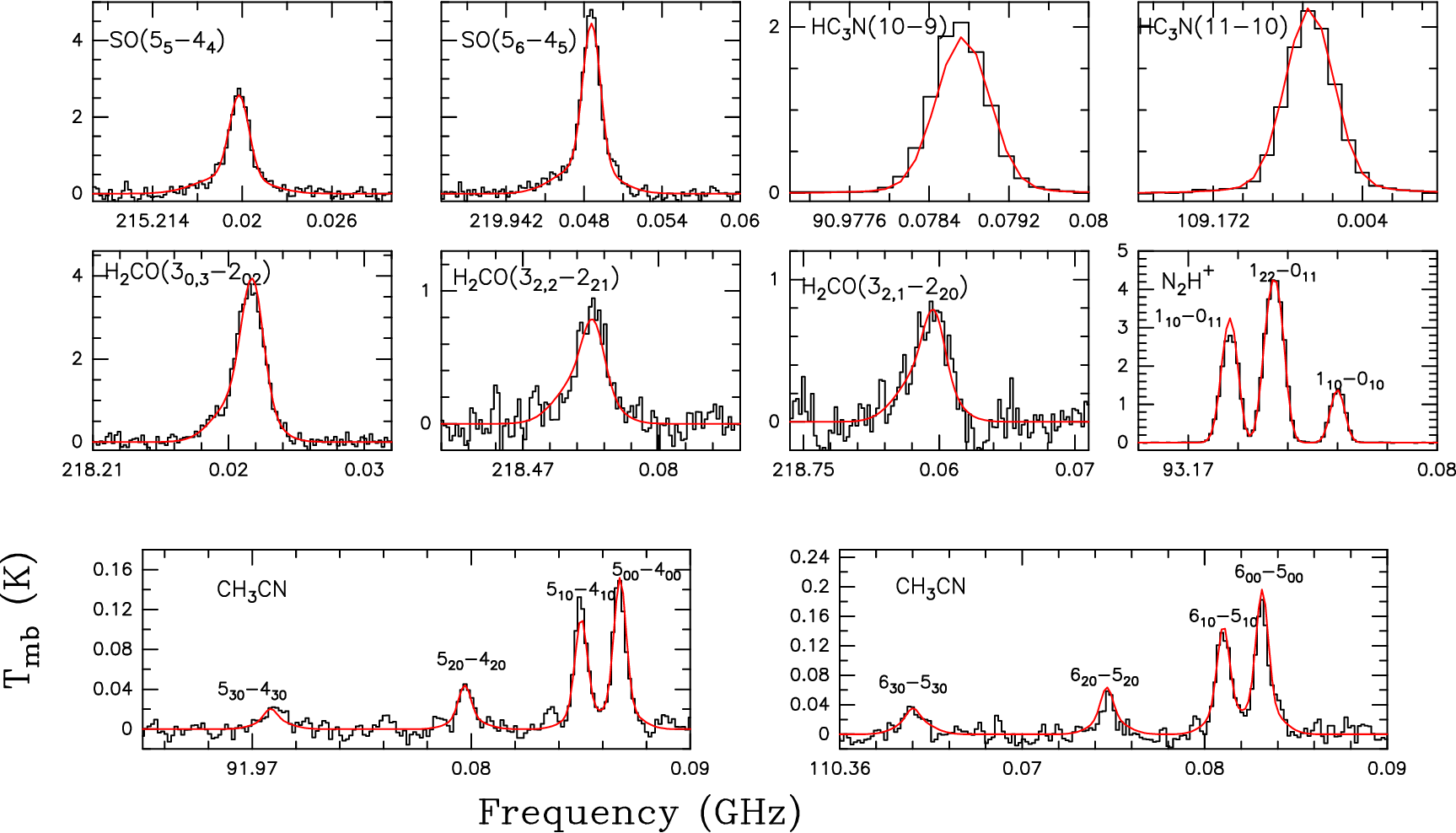}
    \caption{Results of LTE-modeling of spectra for selected transitions of \chthrcn, \htwoco, \hcthn\ and SO at S1. The black histograms show the observed spectra and the red continuous lines correspond to the spectra predicted by the best fit model. All transitions of a particular species have been fitted simultaneously using CASSIS as is mentioned in the text. The reduced $\chi^2$ values for the model fits are 1.68 (\chthrcn), 0.95 (\htwoco), 0.95 (\hcthn), 0.87 (SO), 1.02 (\ntwhp).}
    \label{fig_cassisfit}
\end{figure*}

We fitted selected spectral lines at the centre of S1 to constrain the source size, velocity, linewidth, column density, and excitation temperature using an LTE-based model. We used the Python-based fitting tool within {\sc CASSIS \footnote{Based on an analysis carried out with the CASSIS software and the CDMS spectroscopic database. CASSIS has been developed by IRAP-UPS/CNRS (https://cassis.irap.omp.eu).}} for this purpose.  CASSIS makes use of the Monte Carlo Markov chain (MCMC) method to explore the space of parameters and to find the best combination ($\chi^2$ minimization) of them to reproduce the line profiles. Among the spectral lines we have detected,  H$_2$CO(3--2), CH$_3$CN(5--4), CH$_3$CN(6--5), and \ntwhp(1--0) have the advantage of either having multiple lines due to the $K$ ladder or hyperfine structure, which can constrain the temperature of the source very well. Thus, we individually fitted the spectra of these species as well as the two transitions (12--11) and (10--9) of \hcthn.  Figure\,\ref{fig_cassisfit} shows a comparison of the observed spectra and the best-fit model spectra and the corresponding parameters are presented in Table\,\ref{tab_cassis}. 

We first attempted to fit the two \chthrcn\ rotational lines each with multiple transitions corresponding to the $K$ ladder and could fit all eight transitions only with two physical components: a compact (1\arcsec) warm component (63\,K) reproducing the spectral broadening and a colder component (25\,K) with a larger size. The size of the cold component was found to be consistent with the extent of \chthrcn\ emission (Fig.\,\ref{fig_g183_90GHz_2}). Since the fits are degenerate to size and column density,  the sizes of the warm and cold components were fixed to 1\arcsec\ and 50\arcsec\ and the fitting of the \chthrcn\ spectra was repeated to derive better constraints on the \chthrcn\ column density.  With the exception of \ntwhp, for all other molecules we find that similar to \chthrcn\ two physical components are required to reproduce the observed spectra. We thus constrained the size of the cold component of all molecules using the observed size of the integrated emission from the clump and kept the size of the warm component as a free parameter.  For \ntwhp\ only the cold component is sufficient to fit the observed spectrum and the derived size of the cold component is consistent with the size of the integrated emission. 
 
\begin{table*}
\caption{Results of LTE fitting of spectra \label{tab_cassis}}
{\small
\begin{tabular}{lcrrrrrrrrr}
\hline \hline
  \multicolumn{1}{c}{Molecule} &
  \multicolumn{5}{c}{Component 1} &
  \multicolumn{5}{c}{Component 2}\\
  \cline{3-5} 
  \cline{8-10}
  & $N_{\rm tot}$ & $T_{\rm ex}$ &  Size &
 \vlsr & FWHM & $N_{\rm tot}$ & $T_{\rm ex}$ &  
Size &\vlsr &FWHM \\  
 &(cm$^{-2}$) &(K) &(\arcsec ) & (km\,s$^{-1}$) &
(km\,s$^{-1}$)&(cm$^{-2}$) &(K) &  (arcsec) &(km\,s$^{-1}$) &
(km\,s$^{-1}$)  \\
\hline
&&&&&&&&&&\\
\chthrcn\, & 1.5$\pm$0.2 (15) & 62.6$\pm$6.4  & 1 &  -9.1$\pm$0.3 & 5.8$\pm$0.5  & 2.5$\pm$0.2 (12) & 25.3$\pm$1.2 & 50 & -9.1$\pm$0.1 & 2.2$\pm$0.1\\
\htwoco  &  2.$8\pm$0.2 (15)    &  68.4$\pm$2.1 & 2.9$\pm$0.1  & -8.4$\pm$0.1  & 5.5$\pm$0.2 & 9.4$\pm$0.5 (13) & 33.7$\pm$1.1      & 27.4$\pm$1.4 &   -9.4$\pm$0.1 &  2.5$\pm$0.13\\
\hcthn  & 1.5.$\pm$0.2 (14)  & 65.5$\pm$3.5 & 3.1$\pm$0.1 & -9.1$\pm$0.1 & 5.4$\pm$0.2 & 3.3$\pm$0.1 (12) & 30.9$\pm$1.5 &  30
&-9.1$\pm$0.1 & 2$\pm$0.1\\
SO  & 1.1.$\pm$0.1 (15) & 65.9$\pm$3.8 & 3.6$\pm$0.3 & -8.2$\pm$0.2  &  6.4$\pm$0.4 & 1.3.$\pm$0.1 (14) & 22.5$\pm$1.7& 40      & -8.8$\pm$0.04  &     2$\pm$0.13\\
\ntwhp\, & \ldots & \ldots & \ldots & \ldots & \ldots & 1.2$\pm$0.1 (14) & 27.6$\pm$4.8 & 32 & -9.2$\pm$0.1 & 2.1$\pm$0.1\\
\hline
\end{tabular}}
\end{table*}

The individually fitted \htwoco, \chthrcn, and SO lines with $K_a$ ladders consistently indicate two gas components both centred around $\sim -9$\,\kms: {\em warm:} with $T_{\rm ex}$=63--68\,K and $\Delta v\sim 5.5$--6.4\,\kms; {\em cold:} with $T_{\rm ex}$=25--34\,K and $\Delta v\sim 2$\,\kms\ (\htwoco\ 2.5\,\kms). The observed spectra of the two transitions of \hcthn\ are also consistent with two similar components. All of these transitions have critical densities exceeding 10$^5$\,\cmcub\ and in the presently fitted models the compact warm component has column densities more than 10--50 times the column densities of the cold component. Intuitively it is possible to justify the outcome of the fitting by considering a hot or warm core surrounded by a colder envelope. For \ntwhp(1--0) with a critical density of $5\times 10^4$\,\cmcub, the emission traces out the entire envelope (from the derived size) and also constrains the temperature to 27.6\,K, which is consistent with the value obtained from the other lines.  The column densities of \chthrcn, \htwoco, \hcthn\ , and SO in the warm component are consistent with the mean values observed in other high mass star-forming regions \citep{Gieser2021}. Typically \chthrcn\ is detected close to ultra-compact (UC) \HII\ regions, so its detection of a large column density in G183 with no radio counterpart suggests that this molecule is excited prior to the  UC \HII\ phase of massive star formation. We have used this preliminary analysis of the spectra to identify and constrain the physical parameters of the molecular gas associated with S1, a more detailed physical and chemical model of S1 will be taken up in a future paper (Mookerjea et al. in prep). 


\section{Discussion}
\subsection{Parsec-scale star-forming filament: Its stability and origin}

G183, consisting of a  4.6\,pc-long velocity-coherent filament located in the outer Galaxy and harbouring the massive YSO S1 provides a unique opportunity to study the velocity fields in such extended star-forming structures. The \CeiO(1--0) emission is optically thin and unlike \thCO(1--0) is not affected by the diffuse emission around the filament, and hence is ideal  for the analysis of the velocity fields. The average column density of the filament and the hub (5$\times 10^{22}$\,\cmsq) is somewhat lower than the typical values observed in the inner Galaxy filaments ($\sim 10^{23}$\,\cmsq).  Although the source was first identified as a HFS and the \CeiO(1--0) emission detects multiple shorter branches, the other filaments are much shorter and fainter. We thus consider the main filament to be the primary mass reservoir for the material accreting to form S1.   The main filament shows transonic velocities ($1<\mathcal{M}<2$) in the southern part and supersonic velocities near the star-forming hub. Estimates of mass per unit length of segments of the filament and a comparison with the respective critical values corresponding to a stable equilibrium between gravity and  thermal and turbulent pressure suggests that most parts of the filament are supercritical. However, the filament shows mainly two star-forming clumps, suggesting a possibly significant role of the support due to the magnetic field. We analysed the second-order velocity structure function to understand the origin of the velocity dispersions in the filament and the relative contribution of local velocity effects and large-scale velocity gradients. (Trans-)sonic velocity dispersions up to length scales of 2\,pc indicate the contribution of the internal velocity structure inherent to the filament and the power-law index suggest their possible origin in turbulent cascade. Supersonic velocity dispersions observed at larger length scales are likely due to ordered motion such as accretion onto S1.

\subsection{Properties of the high-mass protostellar object S1}

\begin{figure}
    \centering
    \includegraphics[width=0.49\textwidth]{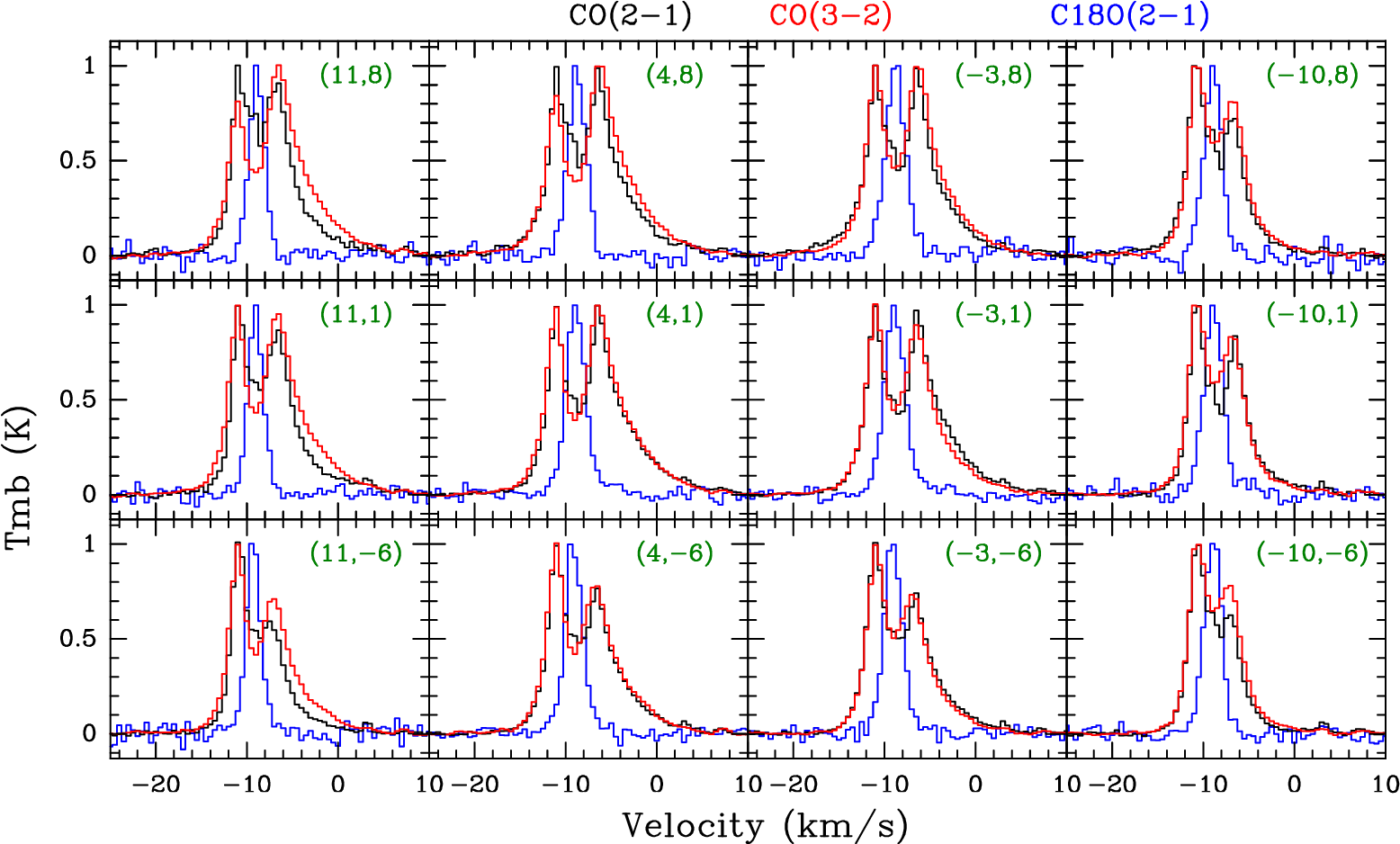}
    \caption{Comparison of normalized spectra of CO(3--2), CO(2--1) and \CeiO(2--1) close to the centre of the hub in G183. All spectra have been convolved to a common angular resolution of 15\arcsec.}
    \label{fig_censpec}
\end{figure}

The region G183 harbours the source IRAS 05480+2545, which was identified as a massive YSO based on the IRAS colour criterion and detection of methanol masers, although no radio continuum emission due to any \HII\ region has been reported to date. More recently, Herschel continuum observations revealed a HFS in the 250\,\micron\ map with the hub coinciding with the IRAS source and estimated a bolometric luminosity of 4170\,\lsun\ for the embedded YSO \citep{Mottram2011}. Non-thermal radio jets have been detected tentatively and the location of the maser sources suggest that their location is in the cavity walls and not in the disc \citep{Purser2021}. Our observations confirm that the most luminous source in G183 coinciding with IRAS 05480+2545 is S1, which shows both signatures of accretion as well as outflow.  We estimate a mass of 156\,\msun\ for S1, which leads to an M/L ratio of 0.04. This value of $M/L$ ratio corresponds to the phase where the YSOs gain mass and increase in luminosity (0.025$\leq M/L\leq$ 0.5\,\msun\,\lsun$^{-1}$), as is observed in a sample of high-mass protostellar objects \citep[HMPOs;][]{Sridharan2002} and the TOP100 ATLASGAL clumps \citep{giannetti_2017}. Thus we propose that the embedded source S1 is an HMPO that currently has a bolometric luminosity consistent with a B1.5 zero-age main-sequence star \citep{Thompson1984}. For a large sample of high-mass star-forming cores, a correlation of the form $M_{\rm out}\sim 0.1 M_{\rm core}^{0.8}$ was derived between the outflow mass $M_{\rm out}$ and core mass ($M_{\rm core}$) \citep{Beuther2002}. Following this correlation, for S1 with an  outflow mass of 2.86\,\msun\ the core mass is expected to be 66\,\msun, which is less than about half the core mass of 156\,\msun\ that we estimated for S1.

Unlike low-mass protostars, for massive YSOs with outflows,  the kinematics at the centre becomes quite complicated and at a resolution of $\sim 0.1$\,pc it is still difficult to disentangle the signatures of rotation and infall from the effects of the large-scale outflow and envelope accretion. A comparison of the CO(2--1), CO(3--2), and \CeiO(2--1) spectra (all at 15\arcsec\ resolution and normalized to 1) close to the centre of the source S1 reveals varying degrees of red and blue asymmetry (Fig.\,\ref{fig_censpec}). We also note that the red and blue asymmetry changes on either side of the centre of S1. This variation in asymmetry of the spectral profiles together with the broadened CO lines suggests an interplay between infall, rotation, and outflow. 

\subsection{On whether G183 is a typical massive star-forming filament}

The recently concluded molecular line surveys of the outer Galaxy have improved the database of such sources \citep{Benedittini2020, Colombo2021}; however, a detailed study of filaments and star formation in the outer Galaxy is still limited. The initially released data of OGHRES survey has identified 37 filaments longer than 10\,pc at a distance of 4 to 10.5\,kpc that have masses and linear masses one order of magnitude lower than similar filaments in the inner Galaxy. Analyses of outer Galaxy filaments and star formation activities therein have resulted in the detection of several long filaments ($>$10\,pc), dense protostellar clumps (10--2000\,\msun), some at the junction of filaments with mass accretion rates between (1--7) 10$^{-4}$ \msun\,yr$^{-1}$ \citep{Guo2022,Sun2023,Rawat2024,Clarke2024}. Based on  observations of CO and its isotopologues in the outer Galaxy filament G214.5-1.8, cloud-scale CO freeze-out due to a low cosmic ray ionization rate was detected \citep{Clarke2024}. Several massive YSOs were identified close to massive central clumps along with H$_2$O maser in the outer Galaxy region between G153.6 and G156.5 \citep{Guo2022}. Together with our detection of hot-core-like chemistry towards G183, in this case associated with a filamentary structure, these results reinforce the view that the processes leading to chemical complexity and massive star formation are robust across the full range of Galactic environments, from the inner Galaxy to the extreme outer disk.

Most of these outer Galaxy filaments including the one in this work show accretion rates along the filaments similar to the inner Galaxy filaments \citep{Mookerjea2023,sen_2024,TrevinoMorales2019}. The outer Galaxy filaments and massive YSOs including G183 show column densities of $<5\times 10^{22}$\,\cmsq, which is on  average lower than the values found in massive YSOs in the inner Galaxy \citep[e.g.][]{TrevinoMorales2019}. The overall properties of G183 are consistent with the few outer Galaxy filaments detected so far. Our work on G183 is the first detailed multi-tracer observation and analysis of a 5-pc long filament with a resolution of 0.3\,pc and its associated massive YSO. Overall, a preliminary comparison of the ratios of line intensities observed in G183 located at a galactocentric distance of 10.44\,kpc with inner Galaxy sources do not show any stark differences. However, as was emphasized earlier, more accurate modelling of the emission along with better constraints on the physical properties that can be obtained from higher-angular-resolution multi-line observations directly comparable to inner Galaxy sources are required.

\section{Summary and outlook}

We have mapped the outer Galaxy source G183 hosting the massive YSO S1 using multiple spectral lines at 1.4 and 3\,mm using the IRAM 30m telescope. The molecular tracers included CO, \thCO, \CeiO, HCN, HNC, \hcop, \ntwhp, \htwoco, \hcthn, SO, and \chthrcn.   The \CeiO(1--0) map tracing the total CO column densities clearly shows a 4.6\,pc long velocity-coherent filament harbouring a massive star-forming hub to the north and another minor clump to the south both with associated far-infrared sources. Using a combination of spectral lines that trace large-scale distribution of molecular gas, high-density tracers, outflow tracers, and hydrocarbons with $K$ ladders, we have obtained a consistent picture of the kinematics in the filament as well as the massive star-forming hub. The properties of the filament derived using the velocity-resolved data are:

\begin{enumerate}[i)]
   \item 
    The width of the \CeiO(1--0) line along the main filament far exceeds the putative thermal line width expected for a  molecular cloud at 20\,K. This implies a significant role of non-thermal contributions (turbulence, magnetic field) in the  broadening of the line. Starting from the south, the filament is transonic ($1<\mathcal{M}<2$) up to 2.4\,pc and thereafter supersonic in the region around S1 and subsequently mildly supersonic further north. We attribute the supersonic velocities to the large-scale velocity variations arising due to accretion and outflow in S1.
    \item 
    The filament, though it is associated with a few mid-IR sources along its length, currently appears to have only two potential star-forming clumps. This is in contradiction with our analysis that the mass per unit length of the filament is supercritical for gravitational collapse considering only thermal and turbulent pressure and indicates additional support; for example, by the magnetic field.
    \item 
    Analysis of the second-order velocity structure along the filament suggests that at small scales, the velocity dispersion is dominated by velocity fields expected from a turbulence cascade, while at large length scales the ordered motion of the star-forming region with accretion and outflow changes the internal velocity significantly.
    \item The intensity-weighted centroid velocity along the length of the filament indicates strong velocity gradients close to S1, and a reversal of the sign of the gradient forming a V-shape consistent with gravitational acceleration of material towards S1. The measured velocity gradients  combined with the mass estimates lead to accretion rates of 8.6\,10$^{-4}$\,\msun\,yr$^{-1}$.
\end{enumerate}

We have confirmed S1 to be a massive YSO and have  constrained its evolutionary stage from its physical properties derived as below:
\begin{enumerate}[i)]
    \item The bolometric luminosity of S1 was known to be 4170\,\lsun. We obtained $M/L = 0.04$ for S1, which is typical of the HMPO stage of evolution.
    \item 
    The parameters estimated for the outflow associated with S1 are consistent with the values found in other high-mass star-forming regions.
    \item 
    Modelling of spectra of complex molecules and high-density tracers at the position of S1 assuming LTE  constrain the emission to be due to a compact hot ($\sim 65$\,K) core and a warm envelope ($\sim$25--30\,K).
\end{enumerate}

A more detailed physical and chemical model of S1 is currently being developed using the maps and this will be combined with our newly obtained higher-angular-resolution multi-transition maps observed with NOEMA and ALMA archival data (Mookerjea et al. in prep). The aim of these analyses will be to constrain the unresolved compact hot component in S1 and also to establish whether the source S1 with no visible UC \HII\ region already harbours a hot core, which would warrant its chemical environment being studied.

\begin{acknowledgements}
B. Mookerjea and S. Sen acknowledge the support of the Department of Atomic Energy, Government of India, under Project Identification No. RTI 4002. This research has made use of the NASA/IPAC Infrared Science Archive, which is funded by the National Aeronautics and Space Administration and operated by the California Institute of Technology.
\end{acknowledgements}

\bibliographystyle{aa}
\bibliography{aa57432-25.bib}{}
\begin{appendix}
\section{Observed molecular lines and their properties }
\begin{table*}
\caption{Properties of the analyzed spectral lines \label{table_1}}
\centering
\begin{tabular}{rcrrccc}
\hline\hline
Molecule & Quantum Numbers &  $\nu^a$ & 
$\log(A_{\mathrm{ul}})^a$& 
$E_\mathrm{u}/k_\mathrm{B}^a$&
$n_{\rm crit}^b$ & 
$n_{\rm eff}^c$\\
& & (GHz) & (log s$^{-1}$) & (K) &(\cmcub) & (\cmcub)\\
\hline
CO & $2-1$ & 230.538 & $-6.16$ & \phn 11.5 & &\\ 
CO & $3-2$ & 345.796 & $-5.60$ & \phn 23.1 & & \\ 
\hline
$^{13}$CO & $1-0$ & 110.201 & $-7.20$ & \phn \phn 5.3 & & \\ 
$^{13}$CO & $2-1$ & 220.399 & $-6.22$ & \phn 11.0 & & \\ 
\hline 
C$^{18}$O & $1-0$ & 109.782 & $-7.20$ & \phn \phn 5.3 & & \\ 
C$^{18}$O & $2-1$ & 219.560 & $-6.22$ & \phn 11.0 & & \\ 
\hline 
SO & $5_{5}-4_{4}$ & 215.220 & $-3.92$ & \phn 34.0 & & \\ 
SO & $5_{6}-4_{5}$ & 219.949 & $-3.87$ & \phn 27.0 & & \\ 
\hline
HCN & $1-0$   & 88.631  & $-4.62$ & \phn \phn 4.3 & 3\,10$^5$ & 4.5\,10$^3$\\
HNC & $1-0$   & 90.663  & $-4.57$ & \phn \phn 4.4 & 1.1\,10$^5$ & 2.3\,10$^3$\\
\hcop & $1-0$ & 89.188  & $-4.38$ & \phn \phn 4.3 & 4.5\,10$^4$ & 5.3\,10$^2$\\
\ntwhp & $1-0$ & 93.173 & $-4.40$ & \phn \phn 4.5 & 4.1\,10$^4$ & 5.5\,10$^3$\\
\hline 
H$_{2}$CO & $3_{0,3}-2_{0,2}$ & 218.222 & $-3.55$ & \phn 21.0 & 7.8\,10$^5$ & 2.0\,10$^5$\\ 
H$_{2}$CO & $3_{2,2}-2_{2,1}$ & 218.476 & $-3.80$ & \phn 68.1 & &\\ 
H$_{2}$CO & $3_{2,1}-2_{2,0}$ & 218.760 & $-3.80$ & \phn 68.1 & & \\ 
\hline 
HC$_{3}$N & $12-11$ & 109.173 & $-3.99$ & \phn 34.1 & 2.1\,10$^5$ & 1.1\,10$^5$\\ 
HC$_{3}$N & $10-9$  & 90.979 & $-4.24$ & \phn 24.0 & 1.2\,10$^5$ & 4.3\,10$^4$\\
\hline
CH$_{3}$CN & $5_{0}-4_{0}$ & 91.987 & $-4.21$ & \phn 13.2 & 1.7\,10$^5$ & 7.4\,10$^4$ \\ 
CH$_{3}$CN & $5_{1}-4_{1}$ & 91.985 & $-4.23$ & \phn 20.4 & & \\ 
CH$_{3}$CN & $5_{2}-4_{2}$ & 91.980 & $-4.29$ & \phn 41.8 & & \\ 
CH$_{3}$CN & $5_{3}-4_{3}$ & 91.971 & $-4.41$ & \phn 77.5 & & \\ 
CH$_{3}$CN & $5_{4}-4_{4}$ & 91.959 & $-4.67$ & \phn 127.5 & & \\ 
CH$_{3}$CN & $6_{0}-5_{0}$ & 110.383 & $-3.96$ & \phn 18.5  & 3.1\,10$^5$ & 1.5\,10$^5$\\ 
CH$_{3}$CN & $6_{1}-5_{1}$ & 110.381 & $-3.98$ & \phn 25.7 & & \\ 
CH$_{3}$CN & $6_{2}-5_{2}$ & 110.374 & $-4.01$ & \phn 47.1 & & \\ 
CH$_{3}$CN & $6_{3}-5_{3}$ & 110.364 & $-4.09$ & \phn 82.8 & & \\ 
CH$_{3}$CN & $6_{4}-5_{4}$ & 110.350 & $-4.21$ & 134.3 & & \\ 
\hline\hline
\end{tabular}
\tablefoot{$^a$ $\nu$, $\log A_{\rm ul}$, $E_{\rm u}/k_{\rm B}$ from Cologne Database for Molecular Spectroscopy \citep[CDMS][]{Mueller2001,Mueller2005,endres2016}, $^b$ $n_{\rm crit}^{\rm thin,nobg}$ for $T_{\rm k}$=20\,K \citep{Shirley2015}, $^c$ $n_{\rm eff}$ for $T_{\rm k}$ = 20\,K \citep{Shirley2015}}
\end{table*}
\section{Dust temperature and column density}
\begin{figure}
    \centering
    \includegraphics[width=0.48\textwidth]{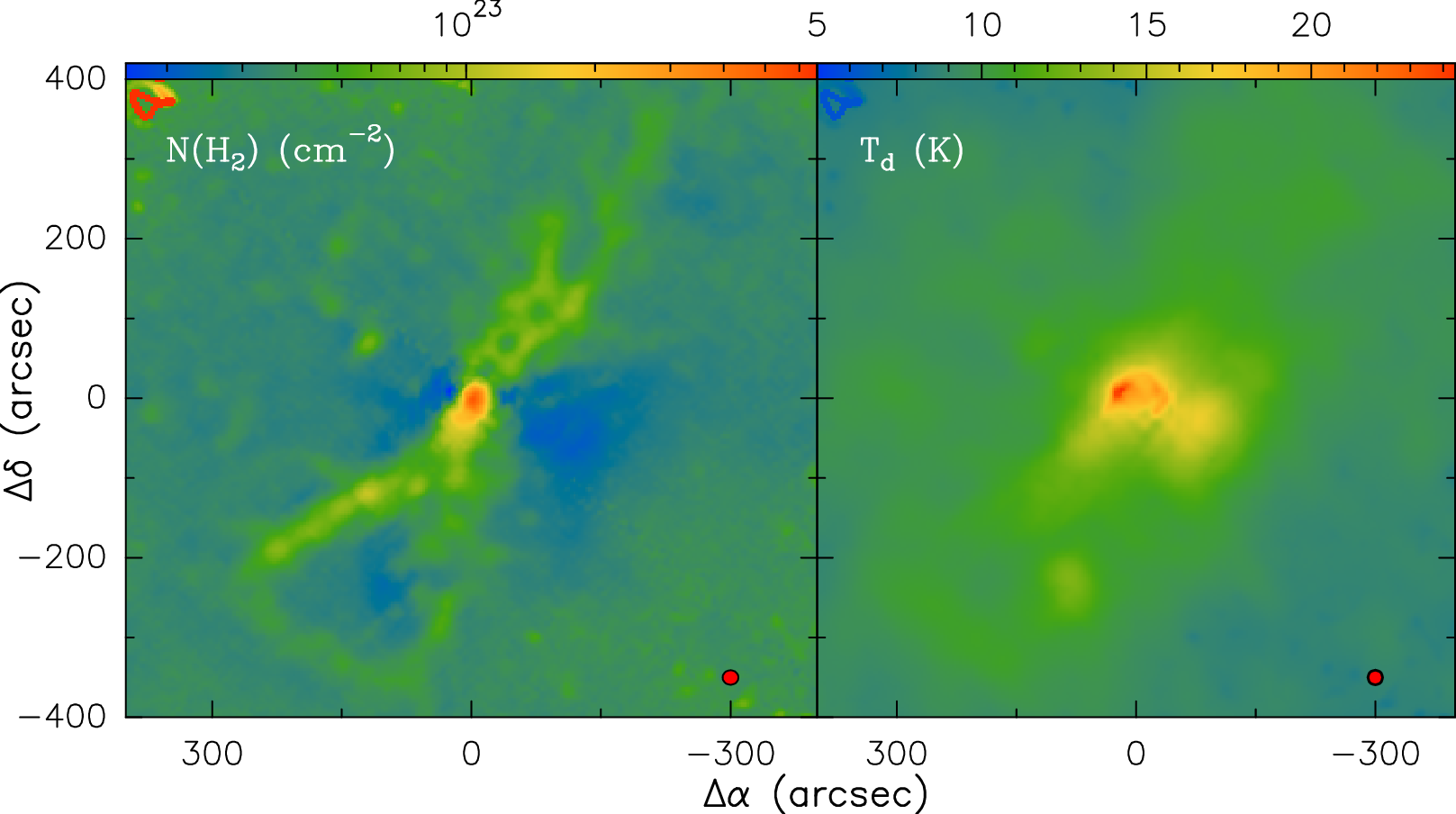}
    \caption{(Left) Colour image of column density ($N$(H$_2$)) in units of cm$^{-2}$ estimated from pixel by pixel greybody fitting of 160, 250, 350 and 500\,\micron\ dust continuum emission. (Right) Colour image of the corresponding fitted dust temperature. The resolution is shown as filled circles in the bottom right corner of both panels.}
    \label{fig_dcolden}
\end{figure}

We performed pixel-by-pixel grey-body fitting of 160, 250, 350 and 500\,\micron\  dust continuum emission using {\em hires}, an improved multi-scale multi-wavelength algorithm for the derivation of high-resolution (18\arcsec) surface densities from multi-wavelength far-infrared Herschel images \citep{Menshchikov2021}. The {\em hires} algorithm is fully described in the appendix A of \citet{palmerim_2013}. A gas-to-dust ratio of 220 appropriate for a galactocentric distance of 10.44\,kpc\ \citep{Gianetti2017b} was used to convert the dust column density to $N$(H$_2$). Figure\,\ref{fig_dcolden} shows the resulting dust temperature and column density maps.

\section{Velocity-Channel maps}

\begin{figure}
    \centering
    \includegraphics[width=0.90\linewidth]{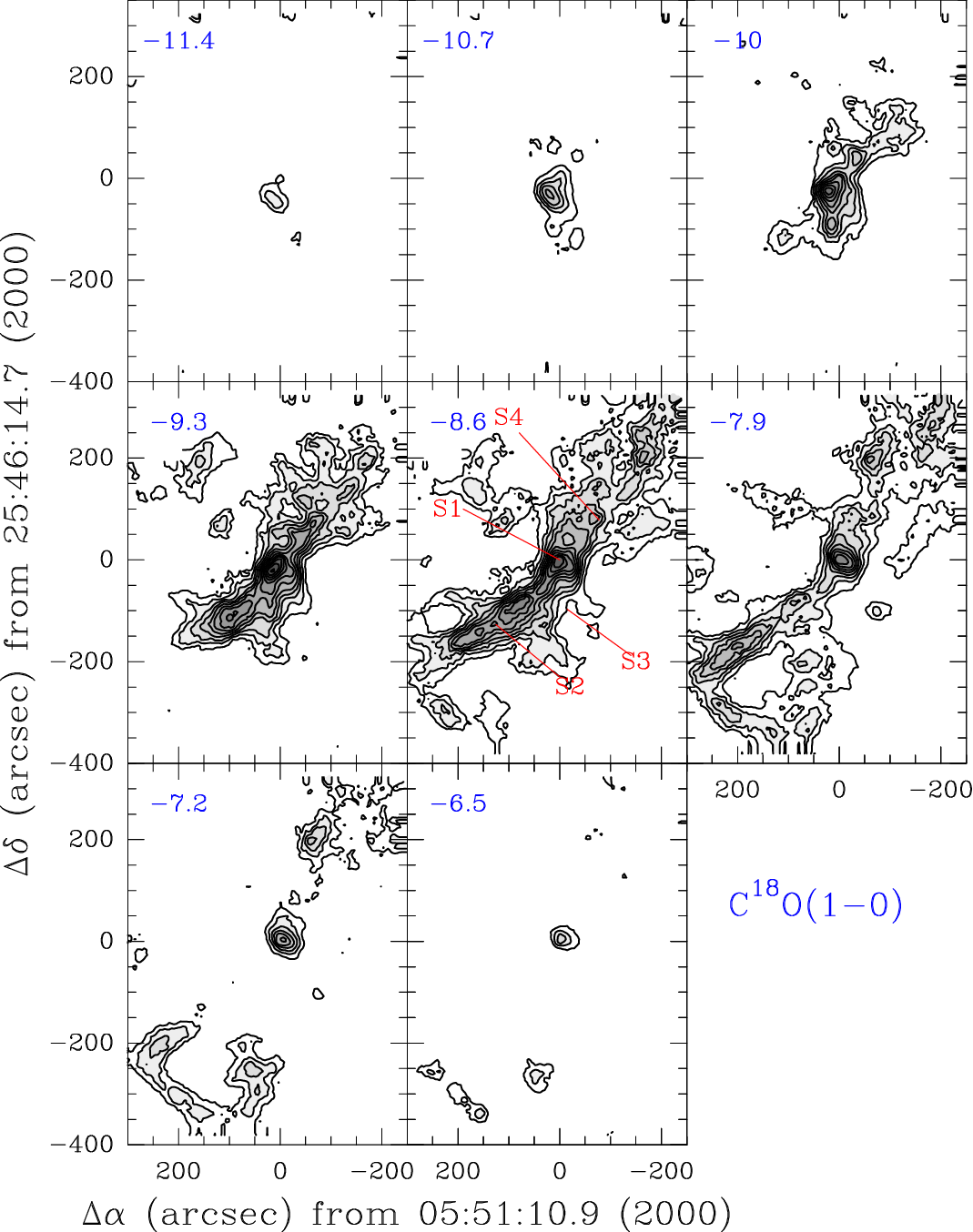}   
    \caption{Velocity channel map \CeiO\,(1--0) with contours at 0.3 to 2.9\,\kms\ in steps of 0.2\,\kms. }
    \label{fig_c18ochanmap}
\end{figure}

\begin{figure}
    \centering
    \includegraphics[width=0.90\linewidth]{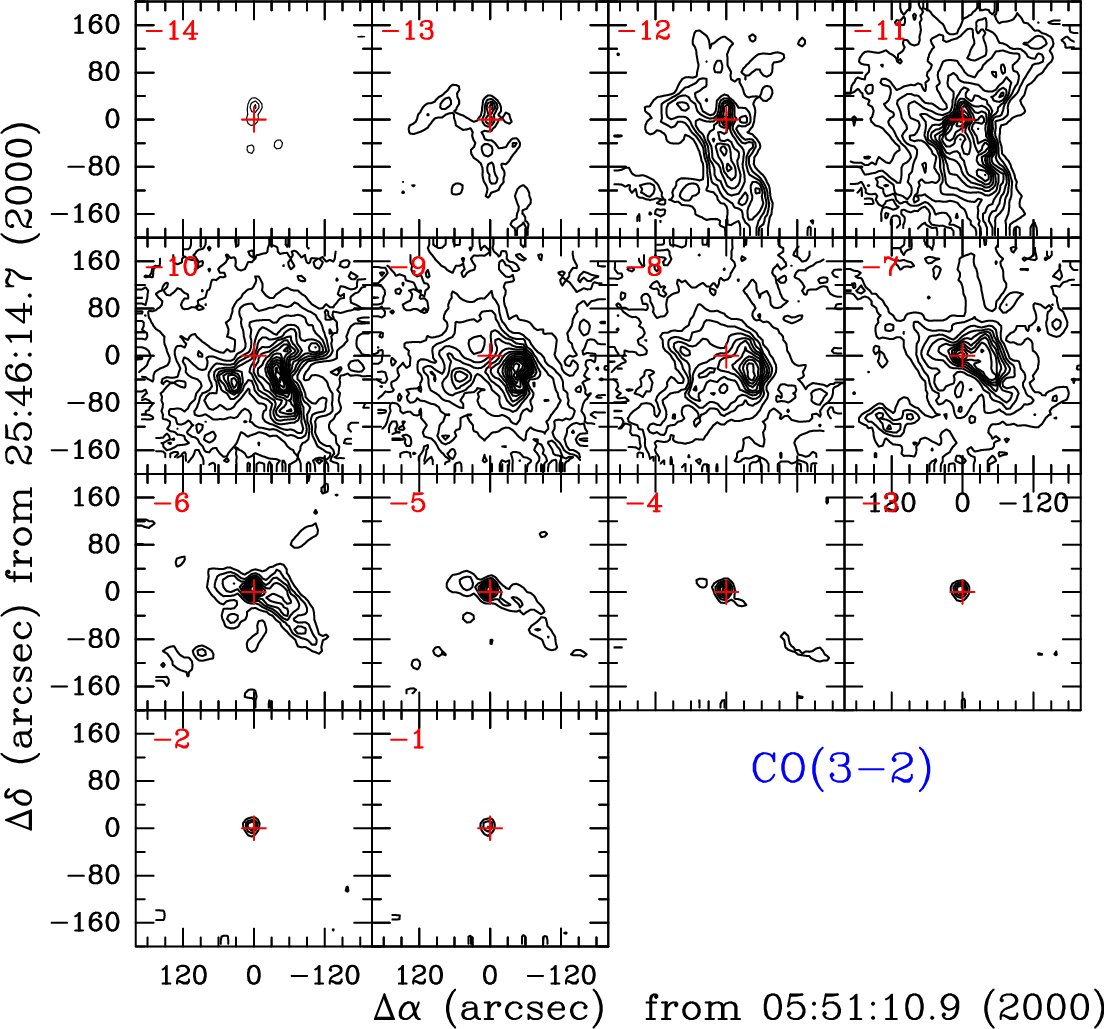}
    \caption{Velocity channel map for CO(3--2) with contours at 1 to 17\,\kms\ in steps of 1\,\kms.
    \label{fig_co32chanmap}}
\end{figure}

\begin{figure}
    \centering
    \includegraphics[width=0.90\linewidth]{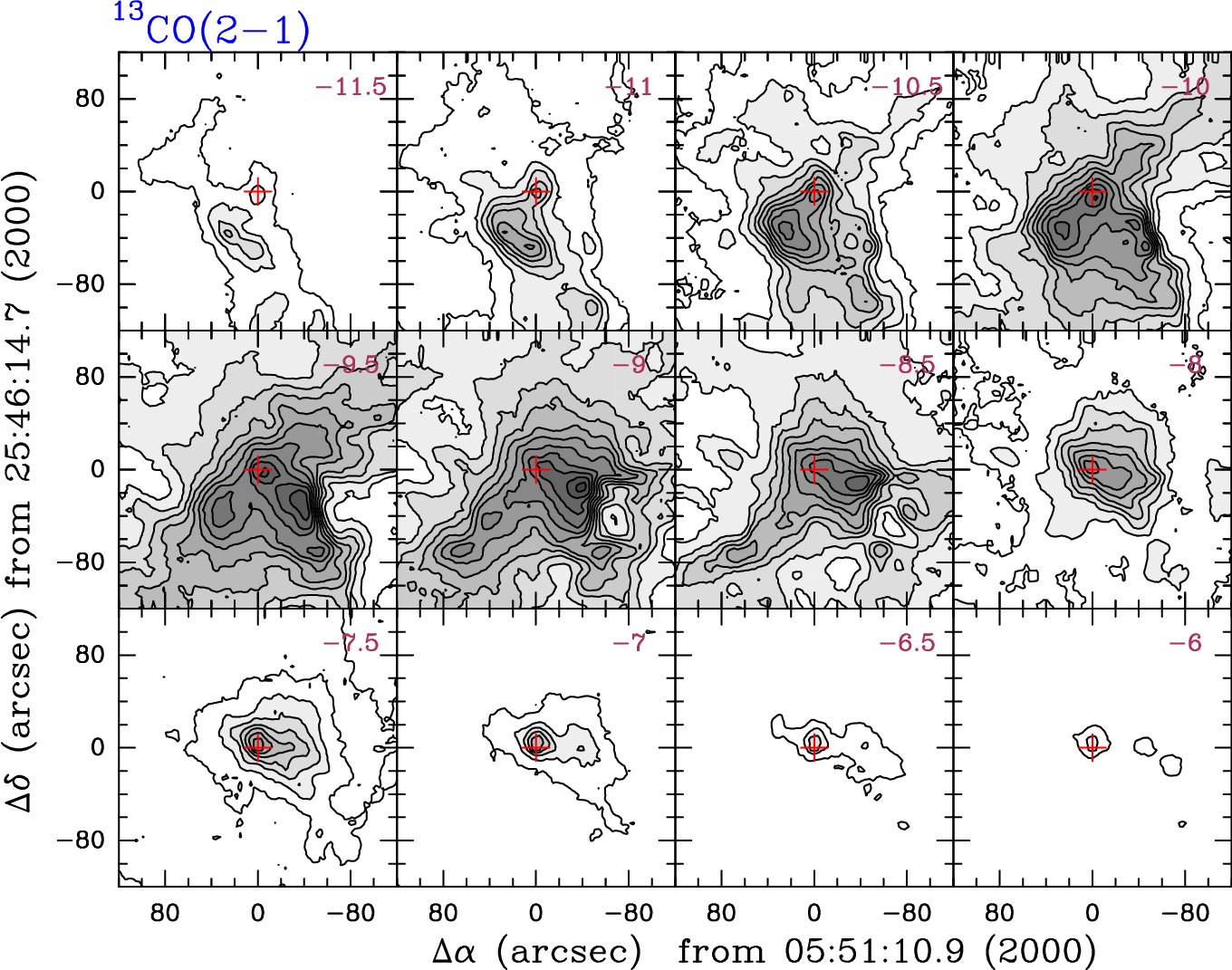}
    \caption{Velocity channel maps for \thCO(2--1) with contours at 1 to 19\,\kms\ in steps of 1.5\,\kms.}
    \label{fig_13co21chanmap}
\end{figure}

\begin{figure}
    \centering    
    \includegraphics[width=0.90\linewidth]{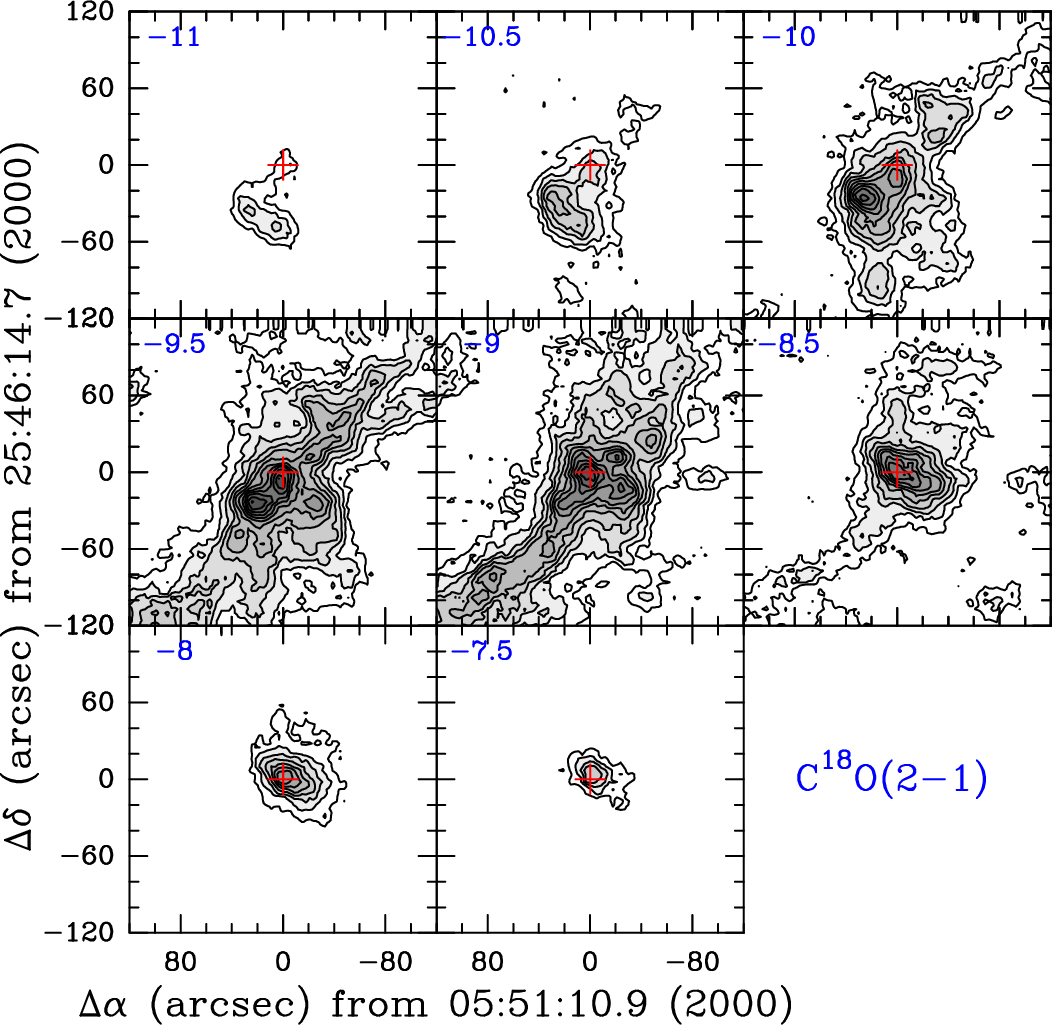}
    \caption{Velocity channel map for \CeiO\,(2--1) with contours at  0.7 to 7.7\,\kms\ in steps of 0.5\,\kms.}
    \label{fig_c18o21chanmap}
\end{figure}

Velocity-channel maps of CO(2--1), CO(3--2) and \thCO(2--1) show that the east-west extended emission is redshifted and could arise partly due to the large-scale outflow driven by the source S1 (Fig.\,\ref{fig_co32chanmap}, \ref{fig_13co21chanmap}). The \CeiO(1--0) and (2--1) channel maps show that the northern part of the long filament is blueshifted while the sickle-shaped southern part is redshifted relative to the observer and also detects outflow close to the central source (Figs.\,\ref{fig_c18ochanmap},\ref{fig_c18o21chanmap}).  Emission from the other dedicated tracers of outflow activities e.g., HCN, \hcop, \htwoco, SO trace the outflow only close to the central condensation. 

\begin{figure}
    \centering
    \includegraphics[width=0.90\linewidth]{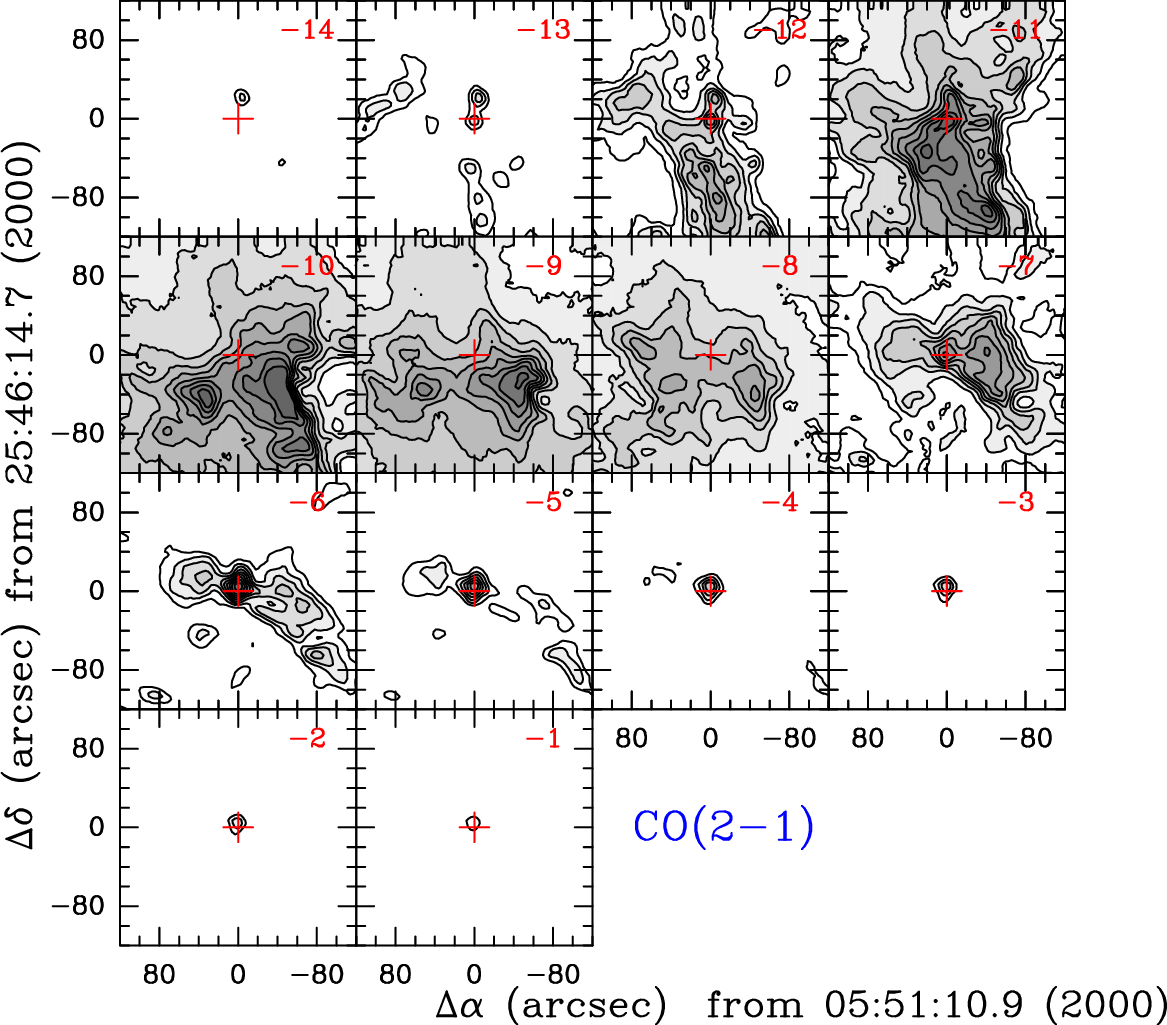}    
    \caption{Velocity channel maps for CO(2--1) with contours at 3 to 21\,\kms\ in steps of 2\,\kms. The red `+' marks the position of the source S1.
    \label{fig_co21chanmap}}
\end{figure}


\section{Definition of outflow parameters \label{sec_outpar}}
We have derived the parameters for the outflow following the formalism developed by \citet{Beuther2002}. The radius ($r_{\rm out}$) of the flow is calculated as he geometric mean of the radii of the red and blue lobes detected in the integrated intensity images of the CO(2--1) emission (Fig.\,\ref{fig_outfmap}). The outflow parameters derived include masses $M_b$, $M_r$ in the blue and the red lobes and the total mass $M_{\rm out}$, the momentum $p$, the energy $E$, the size ($r_{\rm out}$) and the characteristic timescale ($t$, radius of the flow divided by the low velocity), the mass entrainment rate $\dot{M}_{\rm out}$, mechanical force $F_{\rm m}$ and mechanical luminosity $L_{\rm m}$. For mass calculation we used $^{12}$CO/$^{13}$CO = 85.9 and  an excitation temperature of 30\,K, which gives a conversion factor of 8.2$\times 10^{14}$ \cmsq\,K\,\kms\ between the integrated CO(2--1) intensity and column density. 

We use the following relations

\begin{align*} 
M_{\rm out} & = M_{\rm b} + M_{\rm r}\\
p & = M_{\rm b}\times v_{\rm max_b} + M_{\rm r}\times v_{\rm max_r}\\
E & = \dfrac{1}{2}M_{\rm b}\times v_{\rm max_b}^2 + \dfrac{1}{2}M_{\rm r}\times v_{\rm max_r}^2\\
\end{align*} 
\begin{align*} 
t & = \dfrac{r_{\rm out}}{\left(v_{\rm max_b}+v_{\rm max_r}\right)/2}\\
\dot{M}_{\rm out} & = \dfrac{M_{\rm out}}{t}\\
F_{\rm m} & = \dfrac{p}{t}\\
L_{\rm m} & = \dfrac{E}{t}
\end{align*} 
\end{appendix}

\end{document}